\documentclass[a4paper,11pt]{article}

\usepackage{jheppub} 
\usepackage{tikz}

\usepackage{amsmath}
\usepackage{amssymb}
\usepackage{braket}
\usepackage{comment}
\usepackage[many]{tcolorbox}
\usepackage{enumerate}

\newtcolorbox{empheqboxed}{colback=gray!30, 
 colframe=white,
 width=\textwidth,
 sharpish corners,
 top=-2mm, 
 bottom=0mm
}

\newcommand{\TT}{T\overline{T}}

\newcommand*{\Scale}[2][4]{\scalebox{#1}{\ensuremath{#2}}}%

\newcommand{\ul}{\underline}
\newcommand{\U}{\mathcal U}
\newcommand{\V}{\mathcal V}

\newcommand{\C}[1]{$(\ref{#1})$}

\tikzset{
	partial ellipse/.style args={#1:#2:#3}{
		insert path={+ (#1:#3) arc (#1:#2:#3)}
	}
}

\title{Tensionless AdS$_{\boldsymbol 3}$/CFT$_{\boldsymbol 2}$ and Single Trace $\boldsymbol{T\overline{T}}$}

\author[a]{Andrea Dei,}
\author[b]{Bob Knighton,}
\author[c]{Kiarash Naderi,}
\author[a]{Savdeep Sethi}
\affiliation[a]{Enrico Fermi Institute \& Kadanoff Center for Theoretical Physics,\\ \hspace*{0.3cm} University of Chicago, Chicago, IL 60637, USA}
\affiliation[b]{Department of Applied Mathematics \& Theoretical Physics, University of Cambridge,\\
\hspace*{0.3cm}Wilberforce Road, Cambridge CB3 0WA, United Kingdom}
\affiliation[c]{Institut f\"ur Theoretische Physik, ETH Z\"urich,\\ 
\hspace*{0.3cm} Wolfgang-Pauli-Strasse 27, 8093 Z\"urich, Switzerland}
\emailAdd{adei@uchicago.edu}
\emailAdd{rik23@cam.ac.uk}
\emailAdd{knaderi@phys.ethz.ch}
\emailAdd{sethi@uchicago.edu}

\abstract{
One of the few cases of AdS/CFT where both sides of the duality are under good control relates tensionless $k=1$ strings on AdS$_3$ to a two-dimensional symmetric product CFT. Building on prior observations, we propose an exact duality between string theory on a spacetime which is not asymptotically AdS and a non-conformal field theory. The bulk theory is constructed as a marginal deformation of the $k=1$ AdS$_3$ string while the spacetime dual is a single trace $\TT$-deformed symmetric orbifold theory. As evidence for the duality, we match the one-loop bulk and boundary torus partition functions. This correspondence provides a framework to both learn about quantum gravity beyond AdS and understand how to define physical observables in $\TT$-deformed field theories.  
}

\begin{document}
        
\maketitle

\newpage

\section{Introduction and summary}

\subsubsection*{\ul{\it Motivation}} 

There has been striking progress in making the idea of holography precise in the study of three-dimensional gravity with negative cosmological constant. An initial connection between two-dimensional conformal field theory and quantum gravity in asymptotically anti-de Sitter space was first seen in the work of Brown and Henneaux \cite{Brown:1986nw}. The connection was subsequently made more precise in string theory as an example of the AdS/CFT correspondence~\cite{Maldacena:1997re}.

One of the advantages of the special case of AdS$_3$/CFT$_2$ is that it can be realized in string theory with purely NS-NS flux, which admits a conventional worldsheet description. As long as the string coupling is weak, we can use the power of worldsheet string theory to analyze the physics of the bulk AdS solution. The amount $k$ of quantized NS-NS $3$-form flux threading AdS$_3$ characterizes such backgrounds. Large $k$ corresponds to large macroscopic AdS$_3$ spaces.

Our motivation here is to go beyond AdS/CFT and understand the physics of spacetimes that do not behave like a conventional box, which is largely the effect of considering physics in AdS space. We would like to propose a precise holographic correspondence between a non-gravitational field theory and string theory on a spacetime which is not asymptotically AdS. In general this is hard to achieve even in AdS/CFT because one side of the correspondence is typically either strongly-coupled or poorly understood. 

Let us consider the specific example of type II string theory on $\text{AdS}_3 \times \text{S}^3 \times \mathbb T^4 $ with $k$ units of NS-NS flux threading both $\text{AdS}_3$ and $\text S^3$. This background is realized in string theory as the near horizon limit of a collection of $k$ NS5-branes and $n_1$ fundamental strings. The string coupling in this background is fixed in terms of the volume $V_{\mathbb T^4}$, $k$ and the integer $n_1$,
\begin{align}
    g_s^2 =  \frac{V_{\mathbb T^4}}{16\pi^4 (\alpha')^2} \cdot \frac{k}{n_1} \, .
\end{align}
  For fixed $k$, weak string coupling corresponds to large $n_1$.
For this example of AdS$_3$/CFT$_2$ with NS-NS flux and weak string coupling, there is a tractable worldsheet formulation of the bulk string theory; however, the dual spacetime CFT$_2$ is still poorly understood for general $k$. In part, this is because the theory on multiple NS5-branes is also poorly understood. There is a long standing belief that this $\text{AdS}_3$ string background is dual to a CFT$_2$ which is connected to the symmetric product CFT, $\text{Sym}^{N}(\mathbb T^4)$ where $N=k n_1$, by a marginal deformation; for a review see, for example, \cite{David:2002wn}.

Evidence for this belief has accumulated from computations of  \emph{protected} observables in the bulk string theory which do not depend sensitively on the point in moduli space, and therefore can be matched to observables in the symmetric orbifold CFT \cite{Dijkgraaf:1998gf, Larsen:1999uk, Seiberg:1999xz, Argurio:2000tb, deBoer:1998us, Maldacena:1999bp, Gaberdiel:2007vu, Dabholkar:2007ey, Baggio:2012rr, Iguri:2023khc}.
For $k\neq1$, however, the actual spacetime CFT$_2$ is not a symmetric orbifold CFT \cite{Balthazar:2021xeh, Eberhardt:2021vsx}. Moreover, when $k\neq1$ the bulk string spectrum contains a gapped continuum of states that should be reproduced by the boundary CFT$_2$, but is not a part of the symmetric product of $\mathbb T^4$ spectrum. For recent progress and a more in-depth discussion of $\text{AdS}_3$ string theory with $k \neq 1$ see \cite{Eberhardt:2019qcl, Balthazar:2021xeh, Martinec:2021vpk, Eberhardt:2021vsx, Martinec:2022ofs, Dei:2022pkr, Hikida:2023jyc, Knighton:2023mhq, Knighton:2024qxd, Sriprachyakul:2024gyl}. 

On the other hand, the case of $k=1$ is very special \cite{Gaberdiel:2018rqv, Eberhardt:2018ouy}. Restricting to $k=1$ abelianizes the NS5-brane physics, removing all of the mysterious physics supported on multiple NS-branes. This dramatically simplifies the spacetime CFT$_2$. Based on the exact match of the full \emph{non-protected} spectrum, a precise holographic duality between a string-sized $\text{AdS}_3$ and the symmetric orbifold CFT, $\text{Sym}^{N}(\mathbb T^4)$ was proposed a few years ago by Eberhardt, Gaberdiel and Gopakumar~\cite{Eberhardt:2018ouy}: 
\begin{equation}
\begin{tikzpicture}[baseline = -0.6ex]
\node[inner sep=0pt] at (1.5,0.5)
   {\large{Pure NS-NS strings on}};
\node[inner sep=0pt] at (1.5,0)
{$\text{AdS}_3 \times \text{S}^3 \times \mathbb T^4 $}; 
   \node[inner sep=0pt] at (1.5,-0.5)
{with $k=1$}; 
   
\node[inner sep=0pt] at (5.2,-0.1)
   {$\Scale[2]{\iff} $};    

\node[inner sep=0pt] at (7.8,0)
   {{$\text{Sym}^N (\mathbb T^4)$ }};  
\end{tikzpicture}    \, .
\label{ads3/cft2}
\end{equation}
This proposal predicts the holographic match of \emph{any} observable, whether protected or not. While the bulk theory --- frequently dubbed the `tensionless string' --- is still quite intricate, the holographic dual is as nice as one could hope. There have subsequently been many checks and tests of \eqref{ads3/cft2}, which strongly support the proposed duality \cite{Eberhardt:2019ywk, Eberhardt:2020akk, Eberhardt:2020bgq, Dei:2020zui, Gaberdiel:2020ycd, Knighton:2020kuh,  Eberhardt:2021jvj, Gaberdiel:2021njm, Gaberdiel:2021kkp, Knighton:2022ipy, Gaberdiel:2022oeu, Naderi:2022bus, Fiset:2022erp, McStay:2023thk, Dei:2023ivl, Aharony:2024fid, Naderi:2024wqx}. 

\subsubsection*{\ul{\it Beyond AdS}} 

The duality \eqref{ads3/cft2} is our starting point. The other ingredient we need is the rather magical $\TT$ deformation of two-dimensional quantum field theory~\cite{Zamolodchikov:2004ce, Smirnov:2016lqw, Cavaglia:2016oda, Jiang:2019epa}. If the initial undeformed theory is a CFT, this irrelevant deformation breaks conformal invariance. However, if the initial theory is supersymmetric then the deformation can be formulated in a way that preserves manifest $(0,1)$, $(1,1)$, $(0,2)$ and $(2,2)$ supersymmetry~\cite{Baggio:2018rpv, Chang:2018dge, Jiang:2019hux, Chang:2019kiu}. The case of $(4,4)$ or $(0,4)$ supersymmetry with a non-abelian $R$-symmetry is of direct interest to us for these models. We also expect preservation of this larger supersymmetry based on the following argument:  there is a general belief that the global $R$-symmetry will be preserved because $\TT$ deformations at least preserve the energy degeneracies of the undeformed theory. This coupled with viewing such models as special cases of theories with $(2,2)$ or $(0,2)$ supersymmetry implies the preservation of the larger supersymmetry. 

We are interested in what happens when one deforms the symmetric product CFT by a $\TT$ deformation. There are two basic choices for such a deformation. The first possibility is deforming the symmetric orbifold $\text{Sym}^N (\mathbb T^4)$ by the  operator $\det(T)$ where $T$ is the stress tensor of the full theory. At leading order in the deformation parameter, this operator takes the form
\begin{equation}
\left( \sum_{i=1}^{N} T_i \right) \left(\sum_{j=1}^{N} {\bar T_j} \right) \,,
\label{double-trace}
\end{equation}
where $T_i$ is the stress energy tensor of $\mathbb T^4$. This is the conventional double trace $\TT$ deformation. The magic of this irrelevant deformation is that the energy spectrum $E_n(\lambda)$ of the deformed theory on a cylinder of radius $R$ can be determined exactly as a function of the deformation parameter $\lambda$ when the initial theory is a CFT, which has no scale other than the radius $R$. In this case, the flow equation for the energy spectrum can be solved in closed form, 
    \begin{align} \label{CFTenergies}
    E_n ( \lambda ) = \frac{1}{2 \lambda} \left( \sqrt{1 + {4 \lambda E_n} + {4 \lambda^2 P_n^2}}- 1 \right) \, .
\end{align}
Here  $E_n$ is the undeformed energy and $P_n$ is the quantized momentum on the spatial circle. Each of these quantities is made dimensionless with an appropriate power of the radius $R$ along with the deformation parameter $\lambda$. The characteristic square-root form implies a Hagedorn density of high energy states for the `good sign' ($\lambda>0$) of the deformation parameter. For the `bad sign' ($\lambda<0$) most of the energies are complex and how to make sense of the theory is unclear. We will restrict our discussion exclusively to the good sign of the deformation.\footnote{Furthermore, we only consider values of the coupling for which the square root in eq.~\eqref{CFTenergies} is real, namely $\lambda \leq \frac{3}{c}$ for theories with $c=\bar{c}$.} For the good sign and the double trace deformation \C{double-trace}, there is no current well understood holographic or string theory interpretation like AdS/CFT; see, however,~\cite{Guica:2019nzm, Hirano:2020nwq, Kawamoto:2023wzj, Apolo:2023vnm,Blacker:2024rje} for interesting approaches to this question.

The other deformation that is special is the single trace combination of stress-tensors which takes the following form at leading order in $\lambda$, 
\begin{equation}
\left( \sum_{i=1}^{N} T_i{\bar T_i} \right) \,. 
\label{single-trace}
\end{equation}
The operator \C{single-trace} also has conformal dimension $(2,2)$ like \C{double-trace}. It simply $\TT$ deforms each block of the symmetric product with the same energy formula \C{double-trace}. While we do not expect a single trace $\TT$ operator for AdS$_3$/CFT$_2$ with $k>1$, there is top down evidence from string theory for a holographic correspondence in which the spacetime CFT$_2$ is deformed by some operator with properties in common with single trace $\TT$~\cite{Giveon:2017nie, Giveon:2017myj, Asrat:2017tzd}. The bulk string theory background is no longer AdS$_3$ but is instead an asymptotically linear dilaton spacetime. Backgrounds that interpolate between AdS$_3$ and a linear dilaton profile have been studied in past work like~\cite{Forste:1994wp,Israel:2003ry} and more recently in \cite{Martinec:2017ztd, Martinec:2018nco, Martinec:2019wzw, Martinec:2020gkv, Brennan:2020bju, Martinec:2022okx, Dei:2024uyx}. The initial evidence for such a correspondence came from the study of long strings primarily in the $M=0$ BTZ solution~\cite{Giveon:2017myj,Giveon:2017nie,Asrat:2017tzd, Hashimoto:2019wct, Hashimoto:2019hqo, Apolo:2019zai, Chakraborty:2020yka}. The string theory was constructed by deforming the pure NS-NS AdS$_3$ string theory with $k>1$ by the exactly marginal worldsheet operator 
\begin{equation}
    J^+(z) \, \bar J^+(z) \ , 
    \label{Jp-Jp}
\end{equation}
where $J^a(z)$, $a \in \{+, -, 3 \}$ denote $\mathfrak{sl}(2,\mathbb R)_k$ currents in the adjoint representation. Specifically, it was shown in \cite{Giveon:2017myj,Hashimoto:2019wct,Hashimoto:2019hqo} that the long-string spectrum of the deformed background in bosonic string theory on $\text{AdS}_3\times X$ reproduces the spectrum of a single-trace $\TT$-deformed symmetric orbifold on the boundary. See also \cite{Araujo:2018rho,Apolo:2019zai,Apolo:2021wcn,Demise:2021cfx}, where TsT deformations on the string worldsheet were related to $\TT$ deformations of the spacetime theory.

From the spacetime perspective, there is a striking connection between the characteristic square root energy formula \C{CFTenergies} and the form of the supergravity solutions~\cite{Chang:2023kkq, Apolo:2021wcn}. The mass of BTZ black holes also exhibits the same square root form while the asymptotic symmetry group appears to know about a symmetric product structure with a particular choice of boundary conditions~\cite{Georgescu:2022iyx}. All of this suggests a connection with something like a single trace $\TT$ deformation. 
However unless one restricts to protected observables, the study of the holographic dual to $k>1$ AdS$_3$ strings deformed by the current-current operator \eqref{Jp-Jp} inherits all the challenges that $\text{AdS}_3 /\text{CFT}_2$ with $k>1$ already faces prior to any further deformation. Let us summarize the issues: 
\begin{itemize}
    \item The gapped continuum of states characterizing the AdS$_3$ spectrum for $k \neq 1$ should be reproduced by the boundary theory. Therefore one cannot simply work with rational boundary CFTs like the symmetric orbifold of $\mathbb T^4$.
    \item The exact CFT$_2$ dual to $k \neq 1$ AdS$_3$ strings lacks a first principle definition which would allow the computation of observables like generic correlators or the partition function.  
    \item Even before considering any $\TT$ deformation, the CFT$_2$ exactly dual to AdS$_3$ strings \emph{does not} have the structure of a symmetric orbifold. If it is a deformation of the symmetric product by a twist 2 operator, as proposed in~\cite{Balthazar:2021xeh, Eberhardt:2021vsx}, then there would be an interaction between the various copies of $\mathbb T^4$. As a consequence, even defining what one might call the single trace deformation is non-trivial. 
\end{itemize}

\bigskip

In order to avoid these difficulties, here we focus on the exact duality \eqref{ads3/cft2} and consider deformations of the bulk theory by the current-current operator \eqref{Jp-Jp}. In fact, inspired by the $k>1$ bosonic string computation of \cite{Giveon:2017nie}, we show that also at $k=1$ the appropriate current-current worldsheet deformation dual to the boundary field theory $\TT$ operator is given by \eqref{Jp-Jp}. We then compute the torus partition function of the spacetime holographic theory directly from the worldsheet.\footnote{In particular, we consider a specific AdS$_3$ bulk geometry with torus boundary: the so called `cusp geometry.'} We observe that as an effect of the current-current deformation \eqref{Jp-Jp}, the localizing delta function --- a hallmark of the tensionless string \cite{Eberhardt:2019ywk, Eberhardt:2020akk, Dei:2020zui, Knighton:2020kuh, Dei:2023ivl} --- is regularized and smoothened into an exponential. The localizing delta function of tensionless string theory is recovered in the limit $\lambda \to 0$ by means of the distributional identity 
\begin{equation}
\lim_{\lambda\to 0^{+}}\frac{1}{\lambda}\exp\left(-\frac{\pi|x|^2}{\lambda}\right)=\delta^{(2)}(x) \,.
\end{equation}
In Section~\ref{sec:worldsheet} we show that for any choice of (spacetime) spin structure, the string partition function exactly reproduces the dual partition function of the single trace $\TT$-deformed symmetric orbifold of $\mathbb T^4$, which is given in eq.~\eqref{spacetime-final-pf}. Since this is a \emph{non-protected} observable, we are led to conjecture the exact holographic duality:
\vskip 0.1in
\begin{equation}
\begin{tikzpicture}[baseline = -0.6ex]
\node[inner sep=0pt] at (1.5,0.5)
   {\large{$J^+\bar{J}^+$ deformation of}};
\node[inner sep=0pt] at (1.5,0)
{pure NS-NS strings on}; 
   \node[inner sep=0pt] at (1.5,-0.5)
{$\text{AdS}_3 \times \text{S}^3 \times \mathbb T^4 $ with $k=1$}; 
   
\node[inner sep=0pt] at (5.2,-0.1)
   {$\Scale[2]{\iff} $};    

\node[inner sep=0pt] at (8.6,+0.3)
   {{Single trace $\TT$-deformed}};  
\node[inner sep=0pt] at (8.6,-0.3)
   {{$\text{Sym}^N (\mathbb T^4)$ }};  

\end{tikzpicture}    \quad \ . 
\label{ads3/cft2-deformed}
\end{equation}
\vskip 0.2in

As a byproduct of our analysis, taking the limit of vanishing deformation coupling $\lambda \to 0$, we compute the string partition function of the tensionless string for the cusp geometry background. The result confirms the conjecture of \cite{Eberhardt:2020bgq}. The string partition function we derive exactly agrees with the one computed by Eberhardt for thermal $\text{AdS}_3$ or for the conical singularity geometries: the bulk partition function does not depend on the details of the bulk geometry, but just on the geometry of the boundary and its spin structure. 

In Section~\ref{sec:comparison}, based on worldsheet computations and making contact with \cite{Benjamin:2023nts}, we discuss a non-perturbative completion of the $\TT$ partition function formula. From the bulk perspective, this non-perturbative completion includes states with negative winding. The latter have energy that  scales like $\lambda^{-1}$ for small $\lambda$ and hence decouple from the tensionless string spectrum in the limit $\lambda \to 0$. The $\TT$ deformation parameter $\lambda$ acts as a regulator for negative winding states, regularizing the infinite energy they have at $\lambda = 0$. 

\subsubsection*{\ul{\it Future directions}} 

Before we delve into detailed computations, let us list a few future directions suggested by our work and a few issues that 
deserve further investigation. 

\paragraph{The BRST cohomology of the deformed string:} The hybrid formalism has a topologically twisted $\mathcal N=4$ symmetry algebra on the worldsheet. The supercharges have integer conformal dimension and are used to define the cohomology of physical string states. How is the BRST cohomology of physical states defined in presence of the $J^+ \bar J^+$ deformation? A potential strategy to answer this question might go along the following lines: $J^+ \bar J^+$ deformations of a CFT $\mathcal M$ can be recast as a null gauged coset of the rough form $\frac{\mathcal M \times U(1) \times U(1)}{U(1)_L \times U(1)_R}$~\cite{Giveon:2017myj}. After introducing additional ghosts for the gauged currents, it should be possible to deduce the symmetry algebra of the deformed worldsheet theory from the symmetry algebra of the numerator theory of the coset. The techniques developed in \cite{Gaberdiel:2022als} will probably be an important ingredient. Understanding the BRST cohomology should provide a way of probing what goes wrong in the worldsheet theory when $\lambda$ becomes large enough so that the square root of \C{CFTenergies} becomes complex.  

\paragraph{Double trace $\boldsymbol{T\overline{T}}$ and AdS$_{\boldsymbol 3}$:} In this paper we study the tensionless string theory dual to only the \emph{single} trace $\TT$ deformation~\eqref{single-trace} of the symmetric orbifold of $\mathbb T^4$. Can one derive the string theory dual to the good sign double trace deformation of the symmetric product CFT? Similarly, one can also consider combinations of single trace and double trace deformations, even including `bad sign' flows by the single trace deformation followed by a sufficiently large `good sign' flow by the double trace deformation~\cite{Ferko:2022dpg}. Another related direction is to explore the root-$\TT$ proposal of \cite{Babaei-Aghbolagh:2022uij, Ferko:2022cix, Babaei-Aghbolagh:2022leo} and its potential interpretations via coupling to gravity \cite{Babaei-Aghbolagh:2024hti, Tsolakidis:2024wut} or holography~\cite{Ebert:2023tih} using this worldsheet theory as a starting point. 

\paragraph{Correlation functions of $\boldsymbol{T\overline{T}}$-deformed CFTs from the worldsheet:} The worldsheet should provide rules and checks on how to define good physical observables in $\TT$-deformed CFTs.  

\subsubsection*{\ul{\it Organization}} 

The paper is organized as follows: in Section \ref{sec:setup}, we provide a brief introduction to the hybrid formalism for the special case of $k=1$. Section \ref{sec:boundaryfieldtheory} explains how to compute the partition function of the single trace $\TT$-deformed symmetric orbifold theory. This section is purely field theoretic, requiring no background in string theory. In Section~\ref{sec:worldsheet}, after identifying the worldsheet state dual to the single trace $\TT$ deformation, we use the hybrid formalism to compute the partition function directly from worldsheet string theory. Finally, in Section~\ref{sec:comparison} we discuss the matching of string and field theory partition functions. 

\section{The basic setup}
\label{sec:setup}

As a prelude to our subsequent discussion, let us briefly review some ingredients of the tensionless worldsheet string theory. As mentioned in the introduction, we consider superstring theory on $\text{AdS}_3\times \text{S}^3\times \mathbb{T}^4$ with $k=1$ unit of NS-NS flux. This theory is best described in the `hybrid formalism' developed by Berkovits, Vafa and Witten for this class of backgrounds~\cite{Berkovits:1999im}. This formalism has been reviewed several times in recent literature, see e.g.\ \cite{Gaberdiel:2022als,Gerigk:2012lqa,Demulder:2023bux}. Here we briefly sketch the field-content of the superstring in this formalism, which is essential for the worldsheet calculations carried out in Section~\ref{sec:worldsheet}. The worldsheet theory is described using
\begin{equation} \label{eq:different-theories}
\mathfrak{psu}(1,1|2)_1\oplus(\text{a topologically twisted }\mathbb{T}^4)\oplus(\text{ghosts})\,.
\end{equation}
The three factors in \eqref{eq:different-theories} (anti-)commute and form a topologically twisted $\mathcal N=4$ algebra on the worldsheet whose cohomology describes the physical states. Let us discuss each factor in \eqref{eq:different-theories}.

\paragraph{\boldmath The $\mathfrak{psu}(1,1|2)_1$ WZW model:}
This theory has central charge $c=-2$ and admits a free-field realization in terms of symplectic bosons and fermions, see \cite{Gaiotto:2017euk, Eberhardt:2018ouy, Dei:2020zui}. In \cite{Beem:2023dub, Dei:2023ivl} an alternative free-field realization was proposed, which suits better our purpose and will be our starting point. It consists of the following fields,
\begin{itemize}
	\item a $\beta\gamma$-system $(\beta,\gamma)$ with weights $(1,0)$,
	\item two $bc$-systems $(p_i,\theta^i)$ with $i\in\{1,2\}$ and weights $(1,0)$,
\end{itemize}
in addition to their anti-holomorphic copies. The fields satisfy the OPEs
\begin{equation}
    \beta(z)\gamma(w)\sim -\frac{1}{z-w} \,,\qquad p_i(z)\theta^j(w)\sim \frac{\delta^j_i}{z-w} \,,
\end{equation} 
and the associated action reads
\begin{equation}\label{eq:free-field-action}
S=\frac{1}{2\pi}\int\left(\beta\overline{\partial}\gamma+\bar{\beta}\partial\bar{\gamma}+p_i\overline{\partial}\theta^i+\bar{p}_i\partial\bar{\theta}^i\right)\,.
\end{equation}
Intuitively, $\gamma$ is a map from the worldsheet $\Sigma$ to the conformal boundary $X$ of $\text{AdS}_3$, and $\beta$ is a Lagrange multiplier which imposes the condition that $\gamma$ is holomorphic. Also, the $\theta^i$ can be thought of as $\mathcal{N}=2$ supercoordinates on the conformal boundary of $\text{AdS}_3$ \cite{Dei:2023ivl}. For completeness, we write down the free-field realization of the bosonic $\mathfrak{psu}(1,1|2)_1$ currents,
\begin{subequations}
\begin{align} \label{eq:sl2-currents}
J^+& =\beta\,, &  J^3 & =\beta\gamma+\tfrac{1}{2}(p_i\theta^i)\,, &  J^-& =(\beta\gamma)\gamma+(p_i\theta^i)\gamma\,,\\
K^+ & =p_2\theta^1\,, & K^3& =-\tfrac{1}{2}(p_1\theta^1)+\tfrac{1}{2}(p_2\theta^2)\,, & K^-& =p_1\theta^2\,, \label{eq:su2-currents}
\end{align}
while the fermionic currents take the form,
\begin{align}
S^{+++} &=p_2\,, & S^{+-+}& =p_1 \,, & S^{-++} &=-\gamma p_2 \,, & S^{---}&=(\beta\gamma+p_i\theta^i)\theta^2 
 \,, \\
S^{++-}& =\beta\theta^1 \,, & S^{+--} &=-\beta\theta^2\,, &  S^{--+}&=-\gamma p_1 \,, & S^{-+-}& =-(\beta\gamma+p_i\theta^i)\theta^1  \,. \label{eq:psu-super-currents-d}
\end{align}
We follow the conventions of \cite{Dei:2023ivl} for the $\mathfrak{psu}(1,1|2)_1$ (anti-)commutation relations.
\end{subequations}

\paragraph{The topologically twisted $\mathbb{T}^4$:} The $\mathbb{T}^4$ theory appears in our space-time theory as the seed theory $\text{Sym}^N(\mathbb{T}^4)$, and also in the worldsheet theory, where it is topologically twisted. In both cases, this sector is described by $4$ free compact bosons and $4$ free fermions, for which we adopt the following notation:
\begin{itemize}
	\item $4$ bosons $\partial X^j$ and $\partial \bar{X}^j$ with $j\in\{1,2\}$,
	\item $4$ fermions $\psi^{\alpha,j}$ with $\alpha\in\{+,-\}$ and $j\in\{1,2\}$.
\end{itemize}
Similarly for the anti-holomorphic sector. They satisfy the following OPEs
\begin{equation}
	\partial X^i(z) \partial \bar{X}^j(w) \sim \frac{\delta^{ij}}{(z-w)^2} \,,\quad \psi^{\alpha,i}(z) \psi^{\beta,j}(w) \sim \frac{\epsilon^{\alpha\beta}\epsilon^{ji}}{z-w} \,,
\end{equation}
where $\epsilon^{+-}=\epsilon^{12}=1$. In particular, the fermions realize $\mathfrak{su}(2)_1\oplus\mathfrak{su}(2)_1$. Let us denote by $\mathcal{J}$ the Cartan generator of the $\mathfrak{su}(2)_1$ where both $\psi^{+,1}$ and $\psi^{+,2}$ have charge $+\frac{1}{2}$,
\begin{equation} \label{eq:cartan-j}
	\mathcal{J}=-\frac{1}{2} \Big[ \psi^{+,1} \psi^{-,2} - \psi^{+,2} \psi^{-,1}\Big] \,,
\end{equation}
and similarly for the anti-holomorphic part. In our conventions, the topological twist amounts to a shift of the worldsheet stress-tensor $T \mapsto T+\partial \mathcal{J}$, so that the fermions have integer conformal dimensions after the twist. This in particular implies that the central term in the stress-tensor OPE vanishes.

\paragraph{The ghosts:}
The ghost sector contains two bosons $\rho$ and $\sigma$ with non-zero background charges. The scalar $\sigma$ is the bosonization of the reparametrization $bc$-ghost system in string theory with central charge $c(\sigma)=-26$. The scalar $\rho$, which has central charge $c(\rho)=+28$, is less familiar and is a combination of the superconformal ghost and fermions. The ghosts satisfy the following OPEs
\begin{equation}
	\sigma(z) \sigma(w) \sim -\ln{(z-w)} \,, \quad \rho(z)\rho(w)\sim -\ln{(z-w)} \,.
\end{equation}
As expected in a critical string theory, the central charges of the three factors in \eqref{eq:different-theories} add up to $c=0$.

\section{The boundary field theory partition function}
\label{sec:boundaryfieldtheory}

In this section, we compute the torus partition function of the symmetric orbifold of $\mathbb T^4$ after turning on the single trace $\TT$ deformation. Our final result is equation \eqref{spacetime-final-pf}. The partition function can be directly determined from the constraint of modular invariance and the known deformed partition function for the seed theory \cite{Apolo:2023aho}. We will instead follow the strategy of \cite{Hashimoto:2019wct, Hashimoto:2019hqo} and generalize their results to the supersymmetric $\mathbb T^4$ theory by including explicit dependence on the torus spin structure. We also discuss in detail how to define the $\TT$-deformed partition function in the presence of a background field for the $\mathcal R$-charge of the symmetric orbifold theory. We begin in Section \ref{sec:undeformed-seed} by reviewing the partition function of the seed $\mathbb T^4$ theory before any deformation is switched on. As the second step, in Section \ref{sec:deformed-seed} we discuss the $\TT$-deformed seed theory and its associated partition function. Finally, the torus partition function of the single trace $\TT$-deformed symmetric orbifold is computed in Section \ref{sec:symm-orb-deformed-theory}.

\subsection{The undeformed seed theory}
\label{sec:undeformed-seed}

As we discussed in the previous section, the seed CFT consists of four free fermions and four free compact bosons. The partition function of the bosons reads
\begin{equation}
   \frac{\Theta(t)}{\left| \eta(t)  \right|^{8}}   \,,
\end{equation}
where $t$ is the modular parameter of the spacetime torus and $\Theta(t)$ is the Narain Theta function
\begin{equation}
    \Theta(t) = \sum_{(p, \bar p) \in \Gamma_{4,4}} q^{\frac{p^2}{2}} \bar q^{\frac{\bar p^2}{2}} \,, \qquad q = \exp(2 \pi i t) \,.  
\end{equation}
The partition function of the fermions depends on the choice of spin structure and reads
\begin{equation}
\begin{aligned}
    \widetilde{\text R} \text{ sector:} & \quad  \frac{e^{- \frac{\pi}{t_2}u_2^2}}{\left| \eta(t)  \right|^{4}} \left| \vartheta \begin{bmatrix} 0 \\ 0 \end{bmatrix}(\tfrac{u}{2},t) \right|^4 \,, &  \quad  \text{R sector:} & \quad \frac{e^{- \frac{\pi}{t_2}u_2^2}}{\left| \eta(t)  \right|^{4}} \left| \vartheta \begin{bmatrix} 0 \\ \tfrac{1}{2} \end{bmatrix}(\tfrac{u}{2},t) \right|^4 \,, \\[0.3cm]   
    \widetilde{\text{NS}} \text{ sector:} & \quad \frac{e^{- \frac{\pi}{t_2}u_2^2}}{\left| \eta(t)  \right|^{4}} \left| \vartheta \begin{bmatrix} \tfrac{1}{2} \\ 0 \end{bmatrix}(\tfrac{u}{2},t) \right|^4  \,, & \quad    \text{NS sector:} & \quad \frac{e^{- \frac{\pi}{t_2}u_2^2}}{\left| \eta(t)  \right|^{4}} \left| \vartheta \begin{bmatrix} \tfrac{1}{2} \\ \tfrac{1}{2} \end{bmatrix}(\tfrac{u}{2},t) \right|^4 \,, 
    \label{fermion-pf-T4}
\end{aligned}
\end{equation}
where $\widetilde{\text R}$ and $\widetilde{\text{NS}}$  denote the R and NS sectors, respectively, with the insertion of the fermion number operator $(-1)^F$. See Appendix~\ref{app:theta} for definitions of the various Theta functions. The chemical potentials $u$ and $\bar u$ are associated, respectively, to the $\mathcal{R}$-symmetry currents, see eq.~\eqref{eq:cartan-j}. Notice the factor $e^{- \frac{\pi}{t_2}u_2^2}$ in eqs.~\eqref{fermion-pf-T4}. While this factor is often omitted, it naturally emerges in the path integral formalism. This fact is well-known in the literature; see Appendix~\ref{app:path-integral-free-fermions} where we review its origin. As we briefly discuss in Appendices~\ref{app:Modular properties} and \ref{app:path-integral-free-fermions}, the inclusion of the factor $e^{- \frac{\pi}{t_2}u_2^2}$ results in simpler modular properties and streamlines the discussion in the following sections. Moreover, since $\TT$ deformations can be defined in the Lagrangian formalism, it is natural to adopt the path integral definition~\eqref{fermion-pf-T4}. 

Assembling the bosonic and fermionic contributions, we obtain the following partition functions: 
\begin{equation}
\begin{aligned}
    Z_{\widetilde{\text R}}^{\mathbb T^4}(u,t) & = \frac{e^{- \frac{\pi}{t_2}u_2^2} \, \Theta(t)}{\left| \eta(t)  \right|^{12}} \left| \vartheta \begin{bmatrix} 0 \\ 0 \end{bmatrix}(\tfrac{u}{2},t) \right|^{4} \,, &  \quad   Z_{\text R}^{\mathbb T^4}(u,t) & = \frac{e^{- \frac{\pi}{t_2}u_2^2} \, \Theta(t)}{\left| \eta(t)  \right|^{12}} \left| \vartheta \begin{bmatrix} 0 \\ \tfrac{1}{2} \end{bmatrix}(\tfrac{u}{2},t) \right|^{4} \,, \\[0.3cm]
    Z_{\widetilde{\text{NS}}}^{\mathbb T^4}(u,t) & = \frac{e^{- \frac{\pi}{t_2}u_2^2} \, \Theta(t)}{\left| \eta(t)  \right|^{12}} \left| \vartheta \begin{bmatrix} \tfrac{1}{2} \\ 0 \end{bmatrix}(\tfrac{u}{2},t) \right|^{4}  \,, & \quad    Z_{\text{NS}}^{\mathbb T^4}(u,t) & = \frac{e^{- \frac{\pi}{t_2}u_2^2} \, \Theta(t)}{\left| \eta(t)  \right|^{12}} \left| \vartheta \begin{bmatrix} \tfrac{1}{2} \\ \tfrac{1}{2} \end{bmatrix}(\tfrac{u}{2},t) \right|^{4} \,. 
\end{aligned}
\label{T4-part-fun-pi}
\end{equation}
Using the identity \eqref{theta-phi-trick}, we can trade the chemical potential $u$ for real spin structure parameters\footnote{Strictly speaking, the generalized values of $\phi,\chi$ do not define a spin structure, but rather a $\text{Spin}^{\mathbb{C}}$ structure. Holographically, they correspond to turning on a nontrivial background gauge field in the bulk of $\text{AdS}_3$.} $\phi$ and $\chi$ and rewrite the partition functions \eqref{T4-part-fun-pi} respectively as 
\begin{equation}
\begin{aligned}
    Z_{\widetilde{\text R}}^{\mathbb T^4}(u,t) & = Z^{\mathbb T^4}\begin{bmatrix} \phi \\ \chi \end{bmatrix}(t|0) \,, \qquad \text{with} \quad \frac{u}{2} = \chi t -\phi \,, \\
    Z_{\text R}^{\mathbb T^4}(u,t) & = Z^{\mathbb T^4}\begin{bmatrix} \phi \\ \chi \end{bmatrix}(t|0) \,, \qquad \text{with} \quad \frac{u}{2} = \chi t -(\phi + \tfrac{1}{2}) \,, \\
    Z_{\widetilde{\text{NS}}}^{\mathbb T^4}(u,t) & =  Z^{\mathbb T^4}\begin{bmatrix} \phi \\ \chi \end{bmatrix}(t|0) \,, \qquad \text{with} \quad \frac{u}{2} = (\chi-\tfrac{1}{2})t -\phi \,, \\
    Z_{\text{NS}}^{\mathbb T^4}(u,t) & =  Z^{\mathbb T^4}\begin{bmatrix} \phi \\ \chi \end{bmatrix}(t|0) \,, \qquad \text{with} \quad \frac{u}{2} = (\chi-\tfrac{1}{2})t -(\phi + \tfrac{1}{2}) \,,
\end{aligned}
\label{z-to-phi-theta}
\end{equation}
where, for arbitrary real numbers $\phi$ and $\chi$, we defined
\begin{equation}
    Z^{\mathbb T^4} \begin{bmatrix} \phi \\ \chi \end{bmatrix}(t|0) = \frac{\Theta(t)}{\left| \eta(t)  \right|^{12}} \left| \vartheta \begin{bmatrix} \chi \\ -\phi \end{bmatrix}(t) \right|^4 \,. 
    \label{Z-sp-T4}
\end{equation}
As we review in Appendix~\ref{app:path-integral-free-fermions}, $\chi$ and $\phi$ can be interpreted as the real spin structure for complex fermions on the torus. Coupling the free fermions to a background gauge field is equivalent to allowing the fermions to have real holonomies around the cycles of the torus. 

Notice that in each sector, the relation \eqref{z-to-phi-theta} of the chemical potential $u$ with the real spin structures $\chi$ and $\phi$ is consistent with their behaviour under modular transformations, see Appendix~\ref{app:Modular properties}. In terms of the compact notation \eqref{Z-sp-T4}, one can study the modular properties of the various partition functions. Let us introduce the following notation\footnote{Note that this is a consistent action of $\text{PSL}(2,\mathbb{Z})$ on the spin structure, see e.g.\ eq.~(9.101e) in \cite{Blumenhagen:2013fgp}.}
\begin{equation}\label{eq:modular-transformations-general}
\gamma=\begin{pmatrix}
        a & b \\
        c & d
    \end{pmatrix} \in \text{PSL}(2,\mathbb{Z}) \,, \qquad \gamma \cdot t \equiv \frac{a t + b}{c t +d} \,, \qquad \begin{bmatrix} \gamma \cdot \begin{matrix} \phi \\ \chi \end{matrix} \end{bmatrix} \equiv \begin{bmatrix} a \phi + b \chi \\ c \phi + d \chi \end{bmatrix} \,.
\end{equation}
As we discuss in Appendix~\ref{app:Modular properties}, the modular property of the partition functions in the R, NS, $\widetilde{\text R}$ and $\widetilde{\text{NS}}$ sectors can be written in the compact form
\begin{equation}
    Z^{\mathbb T^4} \begin{bmatrix} \begin{matrix} \phi \\ \chi \end{matrix} \end{bmatrix}(\gamma \cdot t|0) = Z^{\mathbb T^4} \begin{bmatrix} \gamma^{-1} \cdot \begin{matrix} \phi \\ \chi \end{matrix} \end{bmatrix}(t|0) \, .
    \label{Z-T4-theta-phi-modular}
\end{equation}
In fact, it is suggestive to rewrite eq.~\eqref{Z-T4-theta-phi-modular} as
\begin{equation}
    Z^{\mathbb T^4} \begin{bmatrix} \gamma \cdot \begin{matrix} \phi \\ \chi \end{matrix} \end{bmatrix}(\gamma \cdot t|0) = Z^{\mathbb T^4}\begin{bmatrix} \begin{matrix} \phi \\ \chi \end{matrix} \end{bmatrix} (t|0) \,. 
    \label{T4-und-nice-modular-prop}
\end{equation}
Geometrically, this equation tells us that under a change of basis $(t,1)\to(at+b,ct+d)$ of the lattice which defines the torus, the $\mathbb{T}^4$ partition function should remain invariant. Under this change of basis, the modular parameter transforms to $t\to(at+b)/(ct+d)$, while the spin structures transform via matrix multiplication as in \eqref{eq:modular-transformations-general}. That is, \eqref{T4-und-nice-modular-prop} is the statement that the spacetime partition function is independent of the basis chosen for the spacetime torus.

\subsection{The deformed seed theory}
\label{sec:deformed-seed}

In the previous section and in Appendix~\ref{app:path-integral-free-fermions}, we see that the fermionic contribution to the undeformed $\mathbb T^4$ partition function with a chemical potential for the $\mathcal R$-symmetry current \eqref{eq:cartan-j} can be recast in terms of twisted fermions on the torus, obeying the boundary conditions
\begin{subequations}
\begin{align}
\eta^\pm(\sigma_1 + 2\pi, \sigma_2) & = e^{\mp 2 \pi i \chi} \, \eta^\pm(\sigma_1, \sigma_2)  \,, & \quad \tilde \eta^\pm(\sigma_1 + 2\pi, \sigma_2) & = e^{\mp 2 \pi i \chi} \, \tilde \eta^\pm(\sigma_1, \sigma_2)  \,,  \\
\eta^\pm(\sigma_1, \sigma_2 + 2\pi) & = e^{\mp 2 \pi i \phi} \, \eta^\pm(\sigma_1, \sigma_2)  \,, & \quad \tilde \eta^\pm(\sigma_1, \sigma_2 + 2\pi) & = e^{\mp 2 \pi i \phi} \, \tilde \eta^\pm(\sigma_1, \sigma_2)  \,, \label{eta-boundary-cond}
\end{align}
\end{subequations}
where we parameterized the torus coordinate using $x=\sigma_1 + t \sigma_2$. Let us now consider $\TT$ deforming the $\mathbb{T}^4$ theory. As we are going to see momentarily, the formulation in terms of twisted fermions simplifies the discussion and makes modular properties manifest. Our starting point is the undeformed partition function~\eqref{Z-sp-T4}, 
\begin{equation}
   Z^{\mathbb T^4} \begin{bmatrix} \phi \\ \chi \end{bmatrix}(t|0) = \text{Tr}\left[ g q^{h} \tilde g \bar q^{\tilde h} \right] = \frac{\Theta(t)}{\left| \eta(t)  \right|^{12}} \left| \vartheta \begin{bmatrix} \chi \\ -\phi \end{bmatrix}(t) \right|^4 \,,
   \label{Z-T4-undeformed}
\end{equation}
where $g$ and $\tilde g$ are twist operators that implement the boundary conditions \eqref{eta-boundary-cond} and $h$ and $\tilde h$ denote the Hamiltonian of the undeformed $\mathbb T^4$ theory.

We define the deformed partition function as
\begin{equation}
    Z^{\mathbb{T}^4}\begin{bmatrix} \phi \\ \chi \end{bmatrix} (t|\lambda) \equiv \text{Tr}\left[ g q^{H_\lambda} \tilde g \bar q^{\tilde H_\lambda} \right] \,,
\label{Z-lambda-tr}
\end{equation}
where the energy and momenta on the cylinder read \cite{Zamolodchikov:2004ce}
\begin{subequations}
\begin{align}	
H_\lambda + \tilde H_\lambda  &= \frac{1}{2\lambda} \Big( -1+\sqrt{1+4\lambda (h+\tilde h) +4 \lambda^2 (h-\tilde h)^2} \, \Big) \,, \\[0.2cm]
 H_\lambda - \tilde H_\lambda &= h - \tilde h \,.
\end{align}
\label{eq:energy-formula}%
\end{subequations}
Here $\lambda$ is the dimensionless $\TT$ coupling. Notice that the deformed Hamiltonian depends on $\chi$ through the undeformed Hamiltonian $h + \tilde h$, see eq.~\eqref{h}. The deformed partition function \eqref{Z-lambda-tr} can be rewritten in terms of the undeformed partition function \eqref{Z-T4-undeformed} as \cite{Dubovsky:2018bmo, Cardy:2018sdv, Hashimoto:2019wct, Hashimoto:2019hqo}
\begin{equation} \label{eq:w-step-2}
	Z^{\mathbb{T}^4}\begin{bmatrix} \phi \\ \chi \end{bmatrix}(t|\lambda) = \frac{t_2}{2\lambda} \int_{\mathbb{H}} \frac{\text d^2 \tau} {\tau_2^2} \exp \left( -\frac{\pi |t-\tau|^2}{2\lambda \tau_2} \right) Z^{\mathbb T^4}\begin{bmatrix}
		\phi \\ \chi
	\end{bmatrix}(\tau|0) \ , 
\end{equation}
using the integral representation
\begin{equation}
 e^{-2\sqrt{ab}} = \sqrt{\frac{a}{\pi}}  \int_0^\infty  \frac{\text d\tau_2}{\tau_2^{3/2}} e^{-a/\tau_2 - b\tau_2} \,, \qquad \text{Re}[a]>0 \,, \quad \text{Re}[b]>0 \,. 
\end{equation} 
In eq.~\eqref{eq:w-step-2}, $\mathbb{H}$ denotes the upper-half complex $\tau$ plane. In view of the comparison with the worldsheet computation in Section~\ref{sec:worldsheet}, it is useful to rewrite this formula in terms of an integral over the string fundamental domain $\mathcal F$. For an integrable function $f(\tau)$ on the upper half plane we have
\begin{equation} \label{eq:domain}
	\int_{\mathbb{H}} \frac{\text d^2 \tau}{\tau_2^2} f(\tau) = \sum_{\gamma\in \text{PSL}(2,\mathbb{Z})} \int_{\mathcal{F}} \frac{\text d^2 \tau}{\tau_2^2} f(\gamma \cdot \tau) \,.
\end{equation}
Let us introduce a short-hand notation for the modular transformation of $\lambda$ \cite{Datta:2018thy},
\begin{equation} \label{eq:def-lambda-modular}
	\gamma \cdot_{\tau} \lambda = \frac{\lambda}{|c \tau +d|^2} \,.
\end{equation}
It is easy to show that
\begin{equation} \label{eq:check}
	\frac{|t-\gamma\cdot\tau|^2}{\lambda (\gamma\cdot \tau)_2} = \frac{|\gamma^{-1}\cdot t-\tau|^2}{(\gamma^{-1}\cdot_t \lambda) \tau_2} \,.
\end{equation}
Then using eqs.~\eqref{T4-und-nice-modular-prop}, \eqref{eq:domain} and \eqref{eq:check}, the partition function \eqref{eq:w-step-2} can be rewritten as
\begin{equation} \label{eq:ttbar-seed}
	Z^{\mathbb T^4}\begin{bmatrix} \phi \\ \chi \end{bmatrix}(t|\lambda)=\frac{t_2}{2\lambda} \sum_{\gamma\in \text{PSL}(2,\mathbb{Z})}\int_{\mathcal{F}} \frac{\text d^2 \tau} {\tau_2^2} \exp\left( -\frac{\pi|\gamma \cdot t-\tau|^2}{2(\gamma \cdot_{t} \lambda) \tau_2} \right) Z^{\mathbb T^4}\begin{bmatrix} \gamma \cdot \begin{matrix} \phi \\ \chi \end{matrix}\end{bmatrix}(\tau|0) \ .
\end{equation}
Let us study the modular behaviour of the deformed partition function \eqref{eq:ttbar-seed} under a fixed modular transformation $\omega \in\text{PSL}(2,\mathbb{Z})$. We have
\begin{align}
	Z^{\mathbb T^4}&\begin{bmatrix}\omega \cdot \begin{matrix} \phi \\ \chi \end{matrix}\end{bmatrix}(\omega \cdot t|\omega \cdot_t \lambda) \nonumber \\
 & = \frac{t_2}{2\lambda} \sum_{\gamma\in \text{PSL}(2,\mathbb{Z})}\int_{\mathcal{F}} \frac{\text d^2 \tau} {\tau_2^2} \exp \left( -\frac{\pi|(\gamma \omega)\cdot t-\tau|^2}{2 [(\gamma \omega) \cdot_t \lambda] \tau_2} \right)Z^{\mathbb T^4}\begin{bmatrix} (\gamma \omega) \cdot \begin{matrix} \phi \\ \chi \end{matrix}\end{bmatrix}(\tau|0) \ . 
\end{align}
Relabeling the sum over $\gamma \in \text{PSL}(2, \mathbb Z)$ as a sum over $\gamma' = \gamma \omega \in \text{PSL}(2, \mathbb Z)$, we see that the deformed partition function obeys
\begin{equation} \label{eq:w-modular}
    Z^{\mathbb T^4}\begin{bmatrix}\omega \cdot \begin{matrix} \phi \\ \chi \end{matrix}\end{bmatrix}(\omega \cdot t|\omega \cdot_t \lambda) = Z^{\mathbb T^4}\begin{bmatrix} \phi \\ \chi \end{bmatrix}(t|\lambda) \ . 
\end{equation}
As usual for $\TT$-deformed theories, provided the $\TT$ coupling transforms as in \eqref{eq:def-lambda-modular}, the deformed partition function inherits the same modular properties as the undeformed partition function: compare eqs.~\eqref{T4-und-nice-modular-prop} and \eqref{eq:w-modular}.

\subsection{The symmetric orbifold of the deformed theory}
\label{sec:symm-orb-deformed-theory}

We now turn our attention to the partition function of the symmetric orbifold of the $\TT$-deformed theory. The final result is already known in the literature \cite{Hashimoto:2019wct, Hashimoto:2019hqo, Apolo:2023aho} for bosonic CFTs, and we generalize it here to include non-trivial spin structure. In this section we have in mind the deformed partition function of the symmetric orbifold of $\mathbb{T}^4$, but the derivation will work for any single-trace $\TT$-deformed CFT.

The symmetric orbifold $\text{Sym}^N(\mathbb{T}^4)$ is constructed by tensoring $N$ copies of the $\mathbb{T}^4$ CFT and gauging the $S_N$ symmetry which acts by permuting the various copies. It is convenient to allow $N$ to vary and introduce the `grand canonical' partition function
\begin{equation}
\mathfrak{Z}\begin{bmatrix}\phi \\ \chi\end{bmatrix}(\sigma,t|\lambda)=1 + \sum_{N=1}^{\infty}p^NZ^{\mathbb{T}^4}_N\begin{bmatrix}\phi \\ \chi\end{bmatrix}(t|\lambda)\,,
\end{equation}
where $p=e^{2\pi i\sigma}$ is a fugacity conjugate to $N$ and $Z^{\mathbb{T}^4}_N$ is the single trace $\TT$-deformed partition function of $\text{Sym}^N(\mathbb{T}^4)$. The grand canonical partition function admits a much simpler expansion than any individual partition function for fixed $N$, and can be written in the exponential form \cite{Dijkgraaf:1996xw, Maldacena:1999bp, Knighton:2023xzg}
\begin{equation}\label{eq:grand-canonical}
\mathfrak{Z}\begin{bmatrix}\phi \\ \chi\end{bmatrix}(t|\lambda)=\exp\left(\sum_{\gamma:\Sigma'\to\Sigma}\frac{p^{\text{deg}(\gamma)}}{\text{deg}(\gamma)}Z^{\mathbb{T}^4}\begin{bmatrix} \phi \\ \chi \end{bmatrix} (\Sigma'|\lambda)\right)\,.
\end{equation}
Here, the sum is over all connected covering spaces $\Sigma'$ of the torus $\Sigma$ which we place our theory on, and $\gamma$ denotes the corresponding (holomorphic) covering map. For the case where the seed theory is a CFT (which is not the case once we have applied a $\TT$ deformation), this leads to the well-known DMVV formula expressing symmetric orbifold partition functions in terms of Hecke operators \cite{Dijkgraaf:1996xw, Maldacena:1999bp, Knighton:2023xzg}. 

We can write the torus $\Sigma$ as the quotient of the complex plane by the lattice $\mathbb{Z}\oplus\mathbb{Z}t$. The connected covering spaces $\Sigma'$ are in one-to-one correspondence with sublattices of $\mathbb{Z}\oplus\mathbb{Z}t$. Every such sublattice can be written in the form
\begin{equation}
\text{Span}(at+b,ct+d)
\end{equation}
for integers $a,b,c,d$. That is, the covering space $\Sigma'$ is a torus whose sides are the vectors $at+b$ and $ct+d$ in the complex plane. Two such sublattices are equivalent if they are related by a change of basis, i.e.~by a $\text{PSL}(2,\mathbb{Z})$ matrix. The degree $\text{deg}(\gamma)$ of the covering map is the determinant of the integer matrix
\begin{equation}\label{eq:integer-matrix}
\begin{pmatrix}
a & b\\
c & d
\end{pmatrix}\,.
\end{equation}
In order to compare the $\TT$-deformed partition function with the undeformed CFT partition function, we want to express it as a function of the modular parameter $\tau=(at+b)/(ct+d)$. This is achieved by scaling the coordinates of the covering torus in such a way that the B-cycle has periodicity $1$, i.e.~performing the scaling
\begin{equation}\label{eq:coordinate-scaling}
z\to\frac{z}{ct+d}\,,\qquad\bar{z}\to\frac{\bar{z}}{c\bar{t}+d}\,.
\end{equation}
In a CFT, the partition function would be invariant under this scaling, and so the covering space partition function would just be the seed theory partition function with modulus $(at+b)/(ct+d)$. However, the $\TT$-deformed theory is not conformal, and so the situation is slightly more complicated. Specifically, reinstating appropriate powers of the radius $R$, one sees that the dimensionless deformation parameter $\lambda$ scales like
\begin{equation}
\lambda\to\frac{\lambda}{|ct+d|^2}
\label{lambda-scaling}
\end{equation}
under the coordinate transformation \eqref{eq:coordinate-scaling}. See also \cite{Benjamin:2023nts}. In fact, the scaling dimension of $\lambda$ does not receive quantum corrections as we deform the theory, which is reflected in the preserved modular properties of $\TT$-deformed partition functions \cite{Datta:2018thy, Aharony:2018bad}. Thus, the covering space partition function is given by
\begin{equation}
Z^{\mathbb{T}^4}\begin{bmatrix}a\phi+b\chi \\ c\phi+d\chi\end{bmatrix}\left(\frac{at+b}{ct+d}\bigg|\frac{\lambda}{|ct+d|^2}\right)\,.
\end{equation}
The modified spin structures are easily read off by noting that the fermions pick up the monodromies
\begin{equation}
\begin{aligned}
\eta^{\pm}(z+at+b) & =e^{\mp 2\pi i(a\phi+b\chi)}\eta^{\pm}(z)\,,\\
\eta^{\pm}(z+ct+d) & =e^{\mp 2\pi i(c\phi+d\chi)}\eta^{\pm}(z) \,,
\end{aligned}
\end{equation}
on the covering surface.

As mentioned above, the set of covering spaces can be written as a sum over all such $a,b,c,d$ up to a change of basis. A change of basis acts as the left-multiplication of a $\text{PSL}(2,\mathbb{Z})$ matrix on the integer matrix \eqref{eq:integer-matrix}. Every integer matrix of the form \eqref{eq:integer-matrix} admits a unique decomposition
\begin{equation}
\begin{pmatrix}
a & b\\ c & d
\end{pmatrix}=
\gamma'\circ
\begin{pmatrix}
a' & b'\\
0 & d'
\end{pmatrix}\,,
\label{matrix-decomposition}
\end{equation}
with $\gamma'\in\text{PSL}(2,\mathbb{Z})$ and $a'd'=a d - bc$ and $b'\in \{ 0,\ldots,d'-1 \}$ \cite{Iwaniec1997TopicsIC}. It follows that the set of matrices 
\begin{equation}\label{eq:integer-matrices-reduced}
\begin{pmatrix}
a & b\\
0 & d
\end{pmatrix}\,,\qquad a,d>0\,,\qquad b=0,\ldots,d-1\,,
\end{equation}
are in one-to-one correspondence with the covering spaces appearing in the exponential of \eqref{eq:grand-canonical}.\footnote{In \eqref{eq:integer-matrices-reduced} the further restriction $a,d>0$ with respect to \eqref{matrix-decomposition} avoids counting twice the same covering space. \label{ad-positive}}  Thus, we have
\begin{equation}\label{eq:deformed-hecke-sum}
\mathfrak{Z}\begin{bmatrix}\phi \\ \chi\end{bmatrix}(\sigma,t|\lambda)=\exp\left(\sum_{a,d>0}\sum_{b=0}^{d-1}\frac{p^{ad}}{ad}Z^{\mathbb{T}^4}\begin{bmatrix}a\phi+b\chi \\ d\chi\end{bmatrix}\left(\frac{at+b}{d}\bigg|\frac{\lambda}{d^2}\right)\right)\,.
\end{equation}
We can now use the expression \eqref{eq:ttbar-seed} to write the grand canonical partition function in terms of the (undeformed) seed theory partition function, namely
\begin{multline}
\log\mathfrak{Z} \begin{bmatrix}
	    \phi \\ \chi
	\end{bmatrix}\hspace{-3pt} (\sigma,t|\lambda) = \hspace{0pt} \frac{1}{2}\sum_{\substack{a,b,c,d \in \mathbb Z \\ ad-bc >0 }} p^{ad-bc} \\
 \times \,  \frac{t_2}{2\lambda} \int_{\mathcal{F}} \frac{\text d^2 \tau}{\tau_2^2} \exp{\left(- \frac{\pi |at +b - (ct +d)\tau|^2}{2 \lambda \tau_2}\right)} Z^{\mathbb{T}^4} \hspace{-3pt}
 \begin{bmatrix}
	   \begin{matrix}
	        a \phi + b \chi \\ c \phi + d \chi
	    \end{matrix}
	\end{bmatrix} \hspace{-3pt}(\tau|0) \,. 
\label{spacetime-final-pf-int}
\end{multline}
Making again use of the decomposition \eqref{matrix-decomposition}, one can see that the sum over the set of all integer matrices with positive determinant in the right-hand-side of \eqref{spacetime-final-pf-int} is a combination of the sum over $a,b,d$ in \eqref{eq:deformed-hecke-sum} and the sum over $\text{PSL}(2,\mathbb{Z})$ matrices in \eqref{eq:ttbar-seed}.\footnote{The additional factor of $1/2$ in the right-hand-side comes again from the fact that the sum over $a,b,c,d$ overcounts equivalent covering spaces, i.e.~$(a,b,c,d)$ and $(-a,-b,-c,-d)$ label the same covering space.} We also note that the factor of $w = ad$ in the denominator of the argument of the exponential in \eqref{eq:deformed-hecke-sum} is canceled against the transformation
\begin{equation}
\frac{\gamma\cdot t_2}{2(\gamma\cdot_t\lambda)}=w\frac{t_2}{2\lambda} \,, 
\end{equation}
in the prefactor of the integral kernel in \eqref{eq:ttbar-seed}. Here we have defined the action of $\gamma$ on $\lambda$ as in \eqref{eq:def-lambda-modular}, namely 
\begin{equation}
\gamma\cdot_t\lambda=\frac{\lambda}{|ct+d|^2}\,.
\end{equation}
In the following, it will prove useful to rewrite eq.~\eqref{spacetime-final-pf-int} in the more compact form 
\begin{tcolorbox}[left=0pt,right=0pt,top=0pt,bottom=10pt]
\begin{equation}
\hspace{-5pt} \log\mathfrak{Z} \begin{bmatrix}
	    \phi \\ \chi
	\end{bmatrix}\hspace{-3pt} (t, \sigma|\lambda) =  \sum_{w=1}^{\infty} p^w \hspace{-5pt} \sum_{\gamma \in \overline{M}_w} \hspace{-5pt}\frac{t_2}{4\lambda} \int_{\mathcal{F}} \frac{\text d^2 \tau}{\tau_2^2} \exp{\Big( \hspace{-3pt}- \frac{\pi |\gamma \cdot t - \tau|^2}{2 (\gamma\cdot_t \lambda) \tau_2}\Big)} Z^{\mathbb{T}^4} \hspace{-3pt}
 \begin{bmatrix}
	    \gamma \cdot \begin{matrix}
	        \phi \\ \chi
	    \end{matrix}
	\end{bmatrix} \hspace{-3pt}(\tau|0) \,,
\label{spacetime-final-pf}
\end{equation}
\end{tcolorbox}
\noindent where we have defined by $\overline{M}_w$ the set of integer matrices of the form \eqref{eq:integer-matrix} with $ad-bc=w$. 

\section{The string theory partition function}
\label{sec:worldsheet}

In Section \ref{sec:setup} we briefly described the worldsheet sigma model of string theory on $\text{AdS}_3\times\text{S}^3\times\mathbb{T}^4$ with $k=1$ unit of NS-NS flux. In this section, we compute its partition function on a locally $\text{AdS}_3$ geometry with torus boundary: the so called `cusp' geometry. We demonstrate how to recover from the worldsheet the single-trace $\TT$ deformation described in the preceding pages.

As reviewed in Section \ref{sec:setup}, in the hybrid formalism the worldsheet CFT consists of a $\text{PSU}(1,1|2)$ WZW model, a topologically twisted sigma model on $\mathbb{T}^4$, as well as the usual $(b,c)$ conformal ghost system and a scalar $\rho$ with central charge $c(\rho)=28$, which is unique to the hybrid formalism. The one-loop worldsheet partition function thus factorizes into three parts:
\begin{equation}
Z(t,\tau)=Z_{\text{PSU}}(t,\tau)Z^{\mathbb{T}^4}(\tau)Z_{\text{ghosts}}(\tau)\,,
\end{equation}
where $t$ is the modulus of the boundary torus. Since the $\mathbb{T}^4$ sigma model is topologically twisted, its fermions live in the Ramond sector, and so
\begin{equation}
Z^{\mathbb{T}^4}(\tau)=\frac{\Theta(\tau)}{|\eta(\tau)|^{12}}\left|\vartheta_2(0,\tau)\right|^{4}\,.
\end{equation}
The partition function of the ghost sector is slightly more subtle, but was worked out in \cite{Eberhardt:2018ouy}. The result is that the $b,c$ system effectively removes two bosonic oscillators, while the $\rho$ ghost removes two pairs of topologically twisted fermions. That is
\begin{equation}
Z_{\text{ghosts}}(\tau)=|\eta(\tau)|^{8}\left|\vartheta_2(0,\tau)\right|^{-4}\,.
\end{equation}
Thus, the string partition function is simply
\begin{equation}
Z(t,\tau)=Z_{\text{PSU}}(t,\tau)\frac{\Theta(\tau)}{|\eta(\tau)|^{4}}\,,
\end{equation}
and so the only nontrivial part of the worldsheet calculation is the computation of the $\text{PSU}(1,1|2)$ WZW model partition function.

In the next subsections, after reviewing the calculation of the path integral of the $\text{PSU}(1,1|2)$ sigma model on spaces whose conformal boundaries are tori of modulus $t$, we will show that the current-current deformation \eqref{Jp-Jp} of the $\text{PSU}(1,1|2)$ sigma model provides a string partition function which exactly reproduces the $\TT$ partition function derived in Section~\ref{sec:boundaryfieldtheory}, see eq.~\eqref{spacetime-final-pf}. Since these deformations do not act on the ghost or $\mathbb{T}^4$ sigma models, they are purely deformations of the $\text{PSU}(1,1|2)$ theory, and so the string path integral will still factorize even upon adding the deformation:
\begin{equation}\label{eq:still-factorizes}
Z(t,\tau|\lambda)=Z_{\text{PSU}}(t,\tau|\lambda)\frac{\Theta(\tau)}{|\eta(\tau)|^4}\,.
\end{equation}

\subsection{Identifying the worldsheet deformation}

Before calculating the worldsheet partition function, let us briefly argue for the form of the marginal deformation operator on the worldsheet. In \cite{Giveon:2017nie} it was shown that the worldsheet operator dual to the single trace $\TT$ deformation of symmetric orbifold theories takes the form \eqref{Jp-Jp}. The computation was carried out in the near boundary limit of bosonic $\text{AdS}_3$ string theory with level $k > 1$. In this section, we show that this result also persists in the exact $k=1$ superstring theory. This section uses the machinery of the hybrid formalism and thus is somewhat technical. Given that only the result is important in the context of the rest of the paper, this discussion can be skipped without loss of readability of the other sections.

In the spacetime theory, the deformation is by the integrated single-trace $\TT$ operator
\begin{equation}
\int\mathrm{d}^2x\left[\left(\TT\otimes\textbf{1}\otimes\textbf{1}\otimes\cdots\right)+\left(\textbf{1}\otimes\TT\otimes\textbf{1}\otimes\cdots\right)+\left(\textbf{1}\otimes\textbf{1}\otimes\TT\otimes\cdots\right)+\cdots\right]\,.
\end{equation}
From the point of view of the orbifold structure of the symmetric orbifold, this is the operator in the $w=1$ twisted sector of the symmetric orbifold which is derived from the $\TT$ operator of the seed theory.

In the bulk, the single-trace $\TT$ operator should be dual to an operator in the $w=1$ spectrally-flowed sector of the worldsheet theory. We can construct this operator using the DDF operators of \cite{Naderi:2022bus,Dei:2023ivl}. The worldsheet dual of the $w=1$ ground state in the symmetric orbifold is the combination \cite{Dei:2020zui, Dei:2023ivl}
\begin{equation} \label{eq:world-sheet-betabetabar}
\ket{0}^{w=1}\Longleftrightarrow e^{2\rho+i\sigma+iH}\theta^1\theta^2\bar{\theta}^1\bar{\theta}^2\delta^{(2)}(\gamma-x)\,,
\end{equation}
where $\partial H$ is the Cartan of the $\text{SU}(2)$ $\mathcal R$-symmetry in the $\mathbb{T}^4$ sigma model on the worldsheet. By the state-operator correspondence, the operator $\TT$ is equivalent to the state $ \mathcal{L}_{-2}\overline{\mathcal{L}}_{-2}\ket{0}$ in the seed theory, so that 
\begin{equation}
\ket{\Phi} = \mathcal{L}_{-2}\overline{\mathcal{L}}_{-2}\ket{0}^{w=1}   
    \label{Lm2barLm20}
\end{equation}
is the state associated to the single-trace $\TT$ operator.\footnote{At first one may think that the operator \eqref{Lm2barLm20} is dual to the \emph{double} trace deformation instead of the \emph{single} trace deformation of the boundary CFT. However, it was observed in \cite{Bertle:2020sgd} and confirmed in \cite{Gaberdiel:2021njm, Naderi:2022bus} that in the untwisted sector the worldsheet state \eqref{Lm2barLm20} selects the stress tensor of the seed theory, instead of the stress tensor of the full symmetric orbifold. See the discussion around eq.~(2.12) in \cite{Bertle:2020sgd} and their footnote 3. See also \cite{Troost:2011ud}.} On the worldsheet, the operator \eqref{Lm2barLm20} has been identified in \cite{Gaberdiel:2021njm, Naderi:2022bus} and restricting to the holomorphic sector it reads
\begin{multline}
    \mathcal L_{-2} \ket{0} =  \frac{1}{2} e^{2 \rho + i \sigma + i H} e^{i f_1 - i f_2 + 3 \phi + 2 \kappa} \Bigl( \partial^2(i f_1 - i f_2 + 2\phi + 4 i \kappa) \\
    + 2 \partial(i f_1 - i f_2 + 2 \phi + 4 i \kappa) \partial(\phi + i \kappa) \Bigr) \,,
    \label{TT-dual-bosonized}
\end{multline}
where normal ordering is assumed and we bosonized the free fields introduced in Section~\ref{sec:setup} according to 
\begin{subequations}
\begin{alignat}{2}
    \beta &= e^{\phi+i\kappa} \partial(i\kappa) \,, \qquad \quad &  \gamma & =e^{-\phi-i\kappa} \,, \\
	\theta^1 &= e^{i f_1} \,, \qquad \quad & p_1 & = e^{-i f_1} \,, \\
	\theta^2 &= e^{-i f_2} \,, \qquad \quad & p_2 & = e^{i f_2} \,. 
\end{alignat}   
\label{new-free-fields-bosonization}%
\end{subequations}
Given a physical vertex operator $\Phi(x,z)$ on the worldsheet, one can construct an exactly marginal deformation of the form
\begin{equation}
\int\mathrm{d}^2z\int\mathrm{d}^2x\,\widetilde{G}^-_{-1}\overline{\widetilde{G}}^-_{-1}\Phi(x,z)\,,
\label{intGGPhi}
\end{equation}
where $\widetilde{G}^-$ is one of the worldsheet $\mathcal{N}=4$ generators needed to define the physical state conditions of the hybrid formalism, see e.g.~Appendix D of \cite{Dei:2023ivl}.\footnote{The worldsheet supercurrent $\widetilde{G}^-$ has worldsheet conformal dimension $\Delta(\widetilde{G}^-)=2$, so that the state in \eqref{intGGPhi} has conformal dimension $(1,1)$.} For our purposes, it suffices to know that
\begin{equation}
\widetilde{G}^{-}=e^{-2\rho-i\sigma-iH}p_1p_2\partial\gamma+\cdots =  e^{-2\rho-i\sigma-iH} e^{-i f_1 + i f_2} \partial(e^{-\phi - i \kappa}) + \cdots \,,
\label{Gtildem}
\end{equation}
where the dots denote terms that after acting on the state in \eqref{eq:world-sheet-betabetabar} result in contributions vanishing inside any correlation function, see e.g.~\cite{Dei:2023ivl}. Acting with $\widetilde G^-_{-1}$ on the state \eqref{TT-dual-bosonized} and using that
\begin{equation}
     \delta(\gamma) = e^{\phi} \,, 
\end{equation}
we find 
\begin{align}
   \widetilde{G}^-_{-1} \Phi(0, z) & \sim - \partial \phi \partial( i \kappa) e^{2 \phi + i \kappa} + \frac{1}{2} \partial(i f_1 - i f_2) \partial( i \kappa) e^{2 \phi + i \kappa} \\
    & = \beta \delta(\gamma) + \frac{1}{2} \beta_0 (\partial(i f_1 - i f_2) e^{\phi}) \,, \label{almost beta delta} 
\end{align}
where $\partial$ denotes the derivative with respect to the worldsheet insertion $z$ and $\sim$ means equality up to total derivatives in $z$. Translating at a generic $x$ we obtain 
\begin{equation}
    \widetilde{G}^-_{-1} \Phi(x, z) = e^{x \beta_0} \widetilde{G}^-_{-1} \Phi(0, z) e^{- x \beta_0} = \beta \delta(\gamma - x) +\frac{1}{2} \partial_x\left( \partial(i f_1 - i f_2) \delta(\gamma-x) \right) \,. 
\end{equation}
Finally, reinstating the anti-holomorphic dependence and integrating over $z$ and $x$ we find
\begin{equation}
\int\mathrm{d}^2z\int\mathrm{d}^2x\,\widetilde{G}^-_{-1}\overline{\widetilde{G}}^-_{-1}\Phi(x,z) = \int\mathrm{d}^2z\int\mathrm{d}^2x \, \beta\bar{\beta}\delta^{(2)}(\gamma-x) = \int\mathrm{d}^2z\,\beta\bar{\beta}\,, 
\label{beta-beta-bar}
\end{equation}
which is exactly the deformation~\eqref{Jp-Jp}.

\subsection[Quotients of AdS\texorpdfstring{$_3$}{3}]{\boldmath Quotients of AdS\texorpdfstring{$_3$}{3}}
\label{sec:quotients-AdS3}

We are interested in computing worldsheet partition functions in the $k=1$ theory on locally-$\text{AdS}_3$ manifolds whose boundary is a torus. Let us denote by $\mathbb{H}^3$ the Euclidean version of global $\text{AdS}_3$, also known as hyperbolic space. All locally-$\text{AdS}_3$ manifolds can be written as $\mathbb{H}^3/\Gamma$, where $\Gamma$ is some discrete subgroup of $\text{SL}(2,\mathbb{C})$. Demanding that $\mathbb{H}^3/\Gamma$ has a torus boundary, there are three possible choices of $\Gamma$ \cite{Maloney:2007ud,Eberhardt:2020bgq}:
\begin{enumerate}[\indent a)]

    \item $\Gamma\cong\mathbb{Z}$. We can choose the generator of $\Gamma$ to be of the form
    \begin{equation}
    \begin{pmatrix}
    e^{2\pi it} & 0\\
    0 & e^{-2\pi it}
    \end{pmatrix}\,,
    \end{equation}
    where $\text{Im}(t)>0$. The resulting bulk manifold $\mathbb{H}^3/\Gamma$ is Euclidean thermal $\text{AdS}_3$.

    \item $\Gamma\cong\mathbb{Z}\times\mathbb{Z}_M$. This group has two generators, namely 
    \begin{equation}
    \begin{pmatrix}
    e^{2\pi it/M} & 0\\
    0 & e^{-2\pi it/M}
    \end{pmatrix}\,, \qquad
    \begin{pmatrix}
    e^{2\pi i/M} & 0\\
    0 & e^{-2\pi i/M}
    \end{pmatrix}\,.
    \end{equation}
    The resulting bulk manifold $\mathbb{H}^3/\Gamma$ is not smooth for $M\neq 0$, but is instead thermal $\text{AdS}_3$ with a conical singularity running through the non-contractible cycle. We recover case a) by setting $M=1$.

    \item $\Gamma\cong\mathbb{Z}\oplus\mathbb{Z}$. In this case, the generators are non-diagonal and can be brought into the form
    \begin{equation}
    \begin{pmatrix}
    1 & 1\\
    0 & 1
    \end{pmatrix}\,,\qquad
    \begin{pmatrix}
    1 & t\\
    0 & 1
    \end{pmatrix}\,.
    \end{equation}
    The bulk manifold $\mathbb{H}^3/\Gamma$ is smooth except for a cusp singularity running along the non-contractible cycle, and is thus sometimes called the `cusp' geometry.
    
\end{enumerate}
In all three cases, the boundary of $\mathbb{H}^3/\Gamma$ is a torus of modular parameter $t$, and so string theory on all three backgrounds should be dual to a CFT living on that torus. A strong version of the background independence of string theory would imply that the worldsheet theories described by the cases a), b), and c) should be equivalent at the quantum level. Indeed, the equivalence of cases a) and b) was shown explicitly in the case of the $k=1$ worldsheet theory in \cite{Eberhardt:2020bgq}. There, the author showed that the partition functions of the worldsheet theory on geometries a) and b) are equal. As a byproduct of the analysis carried out in the following pages, we are going to show that also the partition function of the tensionless string cusp geometry (i.e.~case c)) equals the one computed in \cite{Eberhardt:2020bgq} for the geometries a) and b). 

For the purposes of computing worldsheet partition functions, we can model our worldsheet theory on the above geometries by considering the orbifold of the $\text{PSU}(1,1|2)$ model under the group $\Gamma$. Since we know how $\text{SL}(2,\mathbb{R})$ generators act on all of the free fields, this can be achieved by evaluating the worldsheet path integrals with twisted boundary conditions under the action of $\Gamma$.
\begin{enumerate}[\indent a)]

    \item In case a), the orbifold is generated by $e^{2\pi i t J^3_0}$. Thus, we identify $\Phi\sim e^{2\pi i t J^3_0} \Phi$ for all fields $\Phi$ in our theory. Explicitly, we have the identifications
    \begin{equation}\label{eq:ident-a}
    \begin{gathered}
    \gamma\sim e^{2\pi i t} \gamma\,,\qquad \beta\sim e^{-2\pi i t}\beta\,,\\
    \theta^i\sim e^{\pi i t}\theta^i\,,\qquad p_i\sim e^{-\pi i t}p_i\,.
    \end{gathered}
    \end{equation}

    \item In this case, the generators are $e^{2\pi i t J^3_0/M}$ and $e^{2\pi iJ^3_0/M}$. These generators give rise to the set of identifications
    \begin{equation}\label{eq:ident-b}
    \begin{gathered}
    \gamma\sim e^{2\pi i t/M}\gamma\sim e^{2\pi i/M}\gamma\,,\qquad \beta\sim e^{-2\pi i t/M}\beta\sim e^{-2\pi i/M}\beta\,,\\
    \theta^i\sim e^{\pi i t/M}\theta^i\sim e^{\pi i/M}\theta^i\,,\qquad p_i\sim e^{-\pi i t/M}p_i \sim e^{-\pi i/M}p_i\,.
    \end{gathered}
    \end{equation}

    \item Finally, in this case the generators are $e^{J^+_0}$ and $e^{tJ^+_0}$, yielding the identifications\footnote{We will discuss the action of the generators on $p_i$ and $\theta^i$ below.}
    \begin{equation}\label{eq:ident-c}
    \gamma\sim\gamma+1\sim\gamma+t\,.
    \end{equation}

\end{enumerate}
Of these quotients, probably the most familiar is c), which corresponds to the usual procedure of defining a torus as the quotient $\mathbb{C}/(\mathbb{Z}\oplus t\mathbb{Z})$ of the complex plane, while the other two are related to c) by concatenation with the exponential map $\gamma\to\exp(2\pi i\gamma/w)$.

In the following we will compute the $\text{PSU}(1,1|2)$ partition function on the cusp geometry and show that the deformation 
\begin{equation}\label{eq:jpjp-def}
S_{\lambda}=S_0+\lambda\int J^+\bar{J}^+\,,
\end{equation}
reproduces the single trace $\TT$ deformation of the boundary CFT.

\subsection{The cusp geometry partition function}
\label{sec:cusp-geometry}

Let us consider the geometry described in case c). The calculation is similar to that of \cite{Hashimoto:2019hqo,Hashimoto:2019wct}, with the additional subtlety of treating the worldsheet fermions and the associated spin structure.\footnote{We consider here all four spacetime spin structures R, NS, $\widetilde{\text{R}}$ and $\widetilde{\text{NS}}$, together with the inclusion of the chemical potential $u$ for the worldsheet current $K^3_0$. This results in a spin-$\mathbb C$ structure labeled by the theta characteristics $\phi, \chi \in \mathbb R$. Usually, in $\text{AdS}_3$ string theory computations the only spacetime spin structures one considers are the NS and the $\widetilde{\text{NS}}$ sectors, which have a trivial contraction when extended to the bulk of $\text{AdS}_3$. This restriction seems however not necessary for the tensionless string, which localizes on the boundary of $\text{AdS}_3$. }

As mentioned above, the cusp geometry is obtained from the $\text{PSU}(1,1|2)$ model by the identification
\begin{equation}
\gamma\sim\gamma+1\sim\gamma+t\,.
\end{equation}
To calculate the worldsheet partition function in this background, we follow the general strategy of \cite{Eberhardt:2020bgq}, which is to treat the worldsheet theory as an orbifold of the theory on global $\text{AdS}_3$. To this end, we introduce two generators $\mathcal U,\mathcal V$ of the $\mathbb{Z}\oplus\mathbb{Z}$ orbifold which are defined to act on $\gamma$ as
\begin{equation}
 \U :\gamma\to\gamma+t\,,\qquad \V :\gamma\to\gamma+1\,,
\end{equation}
and similarly for the complex conjugate $\bar{\gamma}$. Note that the Lagrange multipliers $\beta,\bar{\beta}$ are invariant under the orbifold group. 

Since the spacetime CFT we are interested in also has nontrivial spin structures, we need to appropriately orbifold the worldsheet fermions as well, since these correspond holographically to the spacetime supersymmetry generators \cite{Dei:2023ivl}. We specifically define the action of the orbifold generators $\mathcal U,\mathcal V$ on the worldsheet fermions as
\begin{equation}
\begin{split}
\mathcal U :p_i\to e^{2\pi i\phi}p_i\,,&\qquad \theta^i\to e^{-2\pi i\phi}\theta^i\,,\\
\mathcal V :p_i\to e^{2\pi i\chi}p_i\,,&\qquad \theta^i\to e^{-2\pi i\chi}\theta^i\,,
\end{split}
\end{equation}
where $(\phi,\chi)$ is the spin structure chosen on the boundary torus in the spacetime CFT.\footnote{We emphasize that in the hybrid formalism, all worldsheet fields have integer spin, and so calculations are independent of the \textit{worldsheet} spin structure. Nevertheless, as the fundamental fields are spacetime spinors, their periodicities are dependent on the \textit{spacetime} spin structure, see \cite{Knighton:2024ybs}.} Finally, the ghosts $\sigma,\rho$ and the worldsheet $\mathbb{T}^4$ are unaffected by the orbifold.

As in any orbifold, the worldsheet torus partition function will be found by summing over all twisted boundary conditions. That is, if $\Phi$ collectively denotes the worldsheet fundamental fields, the path integral in the orbifold theory separates into sectors with twisted boundary conditions
\begin{equation}\label{eq:twisted-bc-general}
\Phi(z+\tau)=\mathcal U^a \, \mathcal V^b \, \Phi(z)\,,\qquad\Phi(z+1)=\mathcal U^c \, \mathcal V^d \, \Phi(z)\,.
\end{equation}
The total worldsheet partition function is given as a sum over all such sectors, so we schematically write\footnote{We are going to discuss momentarily the precise set of integers $a,b,c,d$ over which the sum in eq.~\eqref{orbifold-box} runs.}
\begin{equation}
Z_{\text{ws}}\begin{bmatrix}\phi\\\chi\end{bmatrix}(t,\tau|\lambda)=\sum_{a,b,c,d}
\Bigg(\begin{tikzpicture}[baseline=2.5ex]
\draw[thick] (0,0) -- (1,0) -- (1,1) -- (0,1) -- (0,0);
\node[below] at (0.5,0) {$\mathcal U^c \mathcal V^d$};
\node[left] at (0,0.5) {$\mathcal U^a \mathcal V^b$};
\end{tikzpicture}\,\,\Bigg)\,,
\label{orbifold-box}
\end{equation}
where the box denotes the path integral with the twisted boundary conditions \eqref{eq:twisted-bc-general}.

The worldsheet theory we are interested in is a marginal current-current deformation of the action \eqref{eq:free-field-action}. As we already discussed, for the cusp geometry the appropriate current-current deformation is $J^+\bar{J}^+$. Since $J^+=\beta$ in the free field realization, we consider the action
\begin{equation}
S_{\lambda}=\frac{1}{2\pi}\int\left(\beta\overline{\partial}\gamma+\bar{\beta}\partial\bar{\gamma}+p_i\overline{\partial}\theta^i+\bar{p}_i\partial\bar{\theta}^i-\lambda\beta\bar{\beta}\right)\,.
\end{equation}
As the theory is free, we can integrate out $\beta,\bar{\beta}$ using their classical equations of motion, and describe the resulting theory using the effective action
\begin{equation}
S_{\lambda}[\gamma,p,\theta]=\frac{1}{2\pi\lambda}\int\partial\bar{\gamma}\,\overline{\partial}\gamma+\frac{1}{2\pi}\int(p_i\overline{\partial}\theta^i+\bar{p}_i\partial\bar{\theta}^i)\,.
\end{equation}
In order to compute the worldsheet partition function on the cusp geometry, we need to compute the path integral
\begin{equation}
\int\mathcal{D}(\gamma,\bar{\gamma})\int\mathcal{D}(p_i,\theta^i,\bar{p}_i,\bar{\theta}^i)\,e^{-S_{\lambda}}\,.
\end{equation}
Since the $\gamma,\bar{\gamma}$ action is Gaussian, we can evaluate it explicitly by summing over saddles and multiplying by a one-loop determinant.

In the $(a,b,c,d)$ sector, the saddles in the path integral will be those satisfying the twisted boundary conditions
\begin{equation}\label{eq:twisted-bc-abcd}
\begin{split}
\gamma(z+\tau)=\gamma(z)+at+b\,,&\qquad\gamma(z+1)=\gamma(z)+ct+d\,,\\
p_i(z+\tau)=e^{2\pi i(a\phi+b\chi)}p_i(z)\,,&\qquad \theta^i(z+\tau)=e^{-2\pi i(a\phi+b\chi)}\theta^i(z)\,,\\
p_i(z+1)=e^{2\pi i(c\phi+d\chi)}p_i(z)\,,&\qquad \theta^i(z+1)=e^{-2\pi i(c\phi+d\chi)}\theta^i(z)\,.
\end{split}
\end{equation}
Since the path integral in a given sector factorizes, we can first compute the bosonic ($\gamma$) path integral and multiply the result with the fermionic ($p_i,\theta^i$) path integral.

The $\gamma$ path integral is computed as a sum over classical saddles multiplied by a one-loop determinant. The classical equation of motion is $\partial\overline{\partial}\gamma=0$, which on the worldsheet torus has solutions of the form
\begin{equation}
\gamma(z,\bar{z})=Az+C\bar{z}+\gamma_0\,,
\label{gamma-sol}
\end{equation}
where $\gamma_0$ is a constant.\footnote{Note that the compact unbranched connected covering maps of $\mathbb{T}^2$ are again $\mathbb{T}^2$. By lifting to $\mathbb{C}$, it can be shown that the (anti-)holomorphic covering maps are affine linear functions.} Demanding $\gamma$ to obey the twisted boundary conditions \eqref{eq:twisted-bc-abcd}, we find the constants $A,C$ to be
\begin{equation}
A=\frac{(at+b)-(ct+d)\bar{\tau}}{2i\tau_2}\,,\qquad C=-\frac{(at+b)-(ct+d)\tau}{2i\tau_2}\,.
\end{equation}
Thus, the classical action is given by
\begin{equation}
S_\lambda^{\gamma \bar \gamma} = \frac{1}{2\pi\lambda}\int\mathrm{d}^2z\,\partial\bar{\gamma}\overline{\partial}\gamma=\frac{\pi\left|(at+b)-(ct+d)\tau\right|^2}{2\lambda\tau_2}\,,
\label{S-gamma-bar-gamma}
\end{equation}
where we have used the fact that the area of the worldsheet torus is $4\pi^2\tau_2$. The path integral over $\gamma$ is thus given by the saddle-point approximation
\begin{equation}
\frac{t_2}{4\lambda\tau_2}\frac{1}{|\eta(\tau)|^4}\exp\left(-\frac{\pi\left|(at+b)-(ct+d)\tau\right|^2}{2\lambda\tau_2}\right)\,.
\label{saddle}
\end{equation}
The prefactor $t_2/4\lambda\tau_2$ in the one-loop determinant can be deduced by comparison with the standard result for the compact boson.\footnote{Comparison with the partition function of two free bosons gives $t_2/(2\lambda\tau_2)$. An additional factor of $\frac{1}{2}$ comes from the identification of sectors labeled by $(a,b,c,d)$ with sectors labeled by $(-a,-b,-c,-d)$. See also \cite{Callebaut:2019omt, Benjamin:2023nts} for a discussion on this point. Strictly speaking, the extra factor of $\frac{1}{2}$ should not sit in front of the whole partition sum, since the contribution of the identity is uniquely identified by $(a,b,c,d)=(0,0,0,0)$. In order to lighten the notation, we will not make this distinction explicit in the following.} 

For the fermions, the path integral is simply written in terms of Theta functions as
\begin{equation}
\frac{1}{|\eta(\tau)|^4}\left|\vartheta\begin{bmatrix}c\phi+d\chi \\ -a\phi-b\chi\end{bmatrix}(\tau)\right|^4\,.
\end{equation}
Putting the fermion and boson partition functions together, along with the simple contribution from the $\mathbb{T}^4$ and ghosts (see equation \eqref{eq:still-factorizes}), we obtain the result
\begin{equation}
\begin{tikzpicture}[baseline=2.5ex]
\draw[thick] (0,0) -- (1,0) -- (1,1) -- (0,1) -- (0,0);
\node[below] at (0.5,0) {$\mathcal U^c \mathcal V^d$};
\node[left] at (0,0.5) {$\mathcal U^a \mathcal V^b$};
\end{tikzpicture}=\frac{t_2}{4\lambda\tau_2}\exp\left(-\frac{\pi\left|(at+b)-(ct+d)\tau\right|^2}{2\lambda\tau_2}\right)Z^{\mathbb{T}^4}\begin{bmatrix}a\phi + b\chi \\ c\phi + d\chi\end{bmatrix}(\tau|0) \,.
\end{equation}
The full worldsheet path integral is then found by summing over all integers $a,b,c,d$ and integrating $\tau$ over the fundamental domain. This gives the final result for the string partition function
\begin{equation}
Z\begin{bmatrix}\phi\\\chi\end{bmatrix}(t |\lambda) = \sum_{a,b,c,d}\frac{t_2}{4\lambda}\int_{\mathcal{F}}\frac{\mathrm{d}^2\tau}{\tau_2^2}\exp\left(-\frac{\pi\left|(at+b)-(ct+d)\tau\right|^2}{2\lambda\tau_2}\right)Z^{\mathbb{T}^4}\begin{bmatrix}a\phi + b\chi \\ c\phi + d\chi\end{bmatrix}(\tau|0)\,.
\label{string-partition-function}
\end{equation}
In particular, the contribution of the $w = ad-bc =1$ sector amounts to
\begin{equation}
Z\begin{bmatrix}\phi\\\chi\end{bmatrix}(t |\lambda)\Bigg|_{w=1} = \frac{t_2}{2\lambda}\int_{\mathbb{H}}\frac{\mathrm{d}^2\tau}{\tau_2^2}\exp\left(-\frac{\pi\left|t-\tau\right|^2}{2\lambda\tau_2}\right)Z^{\mathbb{T}^4}\begin{bmatrix}\phi \\ \chi\end{bmatrix}(\tau|0)\,. \\[0.3cm]
\label{string-partition-function-w=1}
\end{equation}

\subsection*{The grand-canonical deformed partition function} 

Now, we would like to compare this to the boundary field theory results for the deformed symmetric orbifold partition function derived in Section~\ref{sec:boundaryfieldtheory}. In order to do this, we need to consider the grand-canonical partition ensemble \cite{Kim:2015gak,Eberhardt:2020bgq} and introduce a chemical potential in the worldsheet theory which is weighted by the determinant $ad-bc$ of the integer matrix in question. The most natural way to do this is to introduce a constant $B$-field
\begin{equation}
B_{\text{gc}}
=-\frac{\sigma}{2t_2}\mathrm{d}\gamma\wedge\mathrm{d}\bar\gamma 
\label{B-grand-canonical}
\end{equation}
to the worldsheet background. Since $B_{\text{gc}}$ is a closed two-form on the target space, its three-form flux $H_{\text{gc}}$ vanishes identically, and so the addition of this $B$-field does not change the target space equations of motion for a consistent string background. As one can directly check using eq.~\eqref{gamma-sol}, the effect on the worldsheet theory is that the action picks up a term
\begin{equation}\label{eq:topological-term}
\frac{1}{2\pi}\int B_{\text{gc}}=-2\pi i \, \sigma \, \text{deg}(\gamma)\,,
\end{equation}
where $\text{deg}(\gamma)$ is the degree of the map $\gamma$. In the $(a,b,c,d)$ winding sector, we have $\text{deg}(\gamma)=ad-bc$. Thus, the effect of including the `topological' term \eqref{eq:topological-term} to the action is to introduce a factor of $p^{ad-bc}$ to the worldsheet partition function in the $(a,b,c,d)$ winding sector, where $p=e^{2\pi i\sigma}$. That is, the string partition function reads
\begin{equation}\label{eq:string-one-loop-final}
\sum_{a,b,c,d}p^{ad-bc}\frac{t_2}{4\lambda}\int_{\mathcal{F}}\frac{\mathrm{d}^2\tau}{\tau_2^2}\exp\left(-\frac{\pi\left|(at+b)-(ct+d)\tau\right|^2}{2\lambda\tau_2}\right)Z^{\mathbb{T}^4}\begin{bmatrix}a\phi + b\chi \\ c\phi + d\chi\end{bmatrix}(\tau|0)\,.
\end{equation}

\subsection*{Higher-genus contributions}

In the preceding pages, we computed only the one-loop partition function of the worldsheet theory. While this provides the tree-level spectrum of the string theory in the $\beta\bar\beta$-deformed background, it is in principle possible that there are higher-order corrections (in $g_s$) to the spectrum which must be taken into account. While such corrections indeed exist for string backgrounds with generic tension, a remarkable feature of the $k=1$ theory on $\text{AdS}_3\times\text{S}^3\times\mathbb{T}^4$ theory is that the worldsheet spectrum on thermal $\text{AdS}_3$ or the cusp geometry is one-loop exact. Let us briefly explain how this works, and argue that the one-loop exactness remains even after turning on the $\beta\bar\beta$-deformation.

The free energy of the $k=1$ string on $\text{AdS}_3\times\text{S}^3\times\mathbb{T}^4$ is defined in analogy to an $\mathcal{N}=4$ topological string \cite{Berkovits:1994vy}. Specifically, there exists a twisted $\mathcal{N}=4$ algebra on the worldsheet which is used to define the BRST cohomology as well as the definition of the string free energy. More concretely, the genus $g$ free energy is given (for $g\geq 2$) by \cite{Berkovits:1999im,Dei:2020zui}
\begin{equation}\label{eq:free-energy}
\mathcal{F}_g=\int_{\mathcal{M}_{g}}\left\langle\prod_{\alpha=1}^{g-1}|(G^-,\mu_{\alpha})|^2\prod_{\alpha=g}^{3g-3}|\widetilde{G}^-,\mu_{\alpha}|^2\left(\int_{\Sigma}|\widetilde{G}^+|^2\right)^{g-1}\int_{\Sigma}|J|^2\right\rangle\,,
\end{equation}
see \cite{Dei:2023ivl} for definitions of the individual generators. Now, the fermions $p_i,\theta^i$ enter this correlator only through the definition of $\widetilde{G}^-$, which is proportional to $p_1 p_2$, see eq.~\eqref{Gtildem}. Thus, the number of, say, $p_1$ zero modes minus the number of $\theta^1$ zero modes is
\begin{equation}
\# p_1-\# \theta^1=2g-2\,.
\end{equation}
However, since $p_1,\theta^1$ have conformal weights $\Delta(p_1)=1$ and $\Delta(\theta^1)=0$, the Riemann-Roch theorem tells us that the free-energy \eqref{eq:free-energy} vanishes unless
\begin{equation}
\# p_1-\# \theta^1=g-1\,.
\end{equation}
Thus, the free energy vanishes unless $2g-2=g-1$, i.e.~$g=1$.\footnote{This argument is special to the case of $\text{AdS}_3$ backgrounds whose boundaries are topologically tori. For backgrounds with curved boundaries, the definition of the fermions $p_i,\theta^i$ must be modified \cite{Knighton:2024ybs}.} Furthermore, this argument holds upon including the $\beta\bar\beta$ deformation in the worldsheet action, since this deformation does not modify the zero-mode counting of the fermions.

\subsection*{The tensionless string and the $\boldsymbol{\lambda \to 0}$ limit}

Finally, let us briefly explain how the results of \cite{Eberhardt:2018ouy,Eberhardt:2020bgq} are recovered in the limit in which we turn off the deformation. The distributional identity 
\begin{equation}
\lim_{\lambda\to 0^{+}}\frac{1}{\lambda}\exp\left(-\frac{\pi|x|^2}{\lambda}\right)=\delta^{(2)}(x) \,,
\label{delta-function-identity}
\end{equation}
which holds for complex numbers $x$, tells us that the integration kernel in the string partition function \eqref{eq:string-one-loop-final} obeys
\begin{equation}
\lim_{\lambda\to 0^+}\frac{t_2}{4\lambda\tau_2^2}\exp\left(-\frac{\pi|(at+b)-(ct+d)\tau|^2}{2\lambda\tau_2}\right)=\frac{t_2}{2 \tau_2}\delta^{(2)}((at+b)-(ct+d)\tau)\,.
\label{delta-function-lambda=0-1}
\end{equation}
That is, the worldsheet moduli space integral localizes onto worldsheets satisfying $\tau=(at+b)/(ct+d)$ for some $a,b,c,d\in\mathbb{Z}$. Demanding $t_2,\tau_2 >0$ (i.e.~both moduli lie in the upper half-plane), the delta function only has support for $ad-bc>0$. These values of the worldsheet moduli are precisely those for which there exists a \textit{holomorphic} covering map $\gamma$ from the worldsheet to the boundary. This localization is a hallmark of the tensionless worldsheet theory before deformation \cite{Eberhardt:2019ywk, Eberhardt:2020akk, Dei:2020zui, Knighton:2020kuh, Dei:2023ivl}. The effect of the worldsheet current-current deformation is, then, to `smooth out' the localization of the moduli space path integral.

\subsection*{The sphere partition function}

Finally, it remains to consider the genus zero contribution to the string partition function, the so called sphere partition function. In particular, we are interested in the sphere partition function for the string theory introduced in \eqref{ads3/cft2-deformed}, obtained by deforming the tensionless $\text{AdS}_3 \times \text S^3 \times \mathbb T^4$ cusp geometry by the worldsheet marginal operator $J^+ \bar J^+$. Due to the residual conformal symmetry group $\text{PSL}(2,\mathbb{C})$ of the unpunctured sphere, the sphere partition function is notoriously difficult to define, let alone calculate.\footnote{See however \cite{Eberhardt:2021ynh} for the computation of the disk partition function for the open string and \cite{Eberhardt:2023lwd} for the worldsheet computation of the sphere partition function for pure NS-NS string theory on global AdS$_3$.}

Let us first discuss the sphere partition function for the $\text{AdS}_3$ quotients listed in Section~\ref{sec:quotients-AdS3} in absence of any deformation. While for $k=1$ we are not aware of a first principle derivation of the sphere partition function, the latter can be computed in classical gravity when $k$ is large. In fact, in the limit of small $\alpha'$, one expects that the sphere partition function reproduces the leading contribution to Einstein-Hilbert gravity: 
\begin{equation}
e^{Z_{\text{sphere}}}=e^{-I_{\text{on-shell}}}\,,
\end{equation}
where $I_{\text{on-shell}}$ is the Einstein-Hilbert action evaluated on the semiclassical background in question.\footnote{Strictly speaking the computation should not be carried out in pure AdS$_3$ gravity, but for the full supergravity background including $B$ field and dilaton. However, we expect this not to affect the final result.} Since AdS has constant negative curvature, one can use the Brown-Henneaux formula \cite{Brown:1986nw} and the fact that $R=-6/L^2_{\text{AdS}_3}$ on-shell to write
\begin{equation}
I_{\text{on-shell}}=-\frac{1}{16\pi G}\int\mathrm{d}^3x\sqrt{g}\left(R+\frac{2}{L_{\text{AdS}}^2}\right)=\frac{\mathtt c}{6\pi}\frac{\text{Vol}(\text{AdS}_3)}{L_{\text{AdS}}^3}\,,
\end{equation}
where $\mathtt c$ is the central charge of the dual CFT. In the case of string theory on $\text{AdS}_3$, the central charge appearing in the Brown-Henneaux formula actually depends on the number of times the worldsheet wraps the boundary of $\text{AdS}_3$ \cite{Giveon:1998ns, Eberhardt:2019qcl}, in our case $ad-bc$. The `central charge' is then equal to $\mathtt c=\mathtt c_0(ad-bc)$, with $\mathtt c_0=6$ being the central charge of the $\mathbb{T}^4$ seed CFT. This leads to the expectation that the sphere partition function should take the form
\begin{equation}
Z_{\text{sphere}}=-\frac{ad-bc}{\pi}\frac{\text{Vol}(\text{AdS}_3)}{L_{\text{AdS}}^3}
\end{equation}
in the $(a,b,c,d)$ winding sector, see \cite{Carlip:1994gc, Eberhardt:2020bgq}. Of course, the volume of $\text{AdS}_3$ is formally infinite, and so one must use the regularized volume (or the regularized EH action), defined through holographic renormalization. The result depends on the choice of the specific locally AdS$_3$ background. As we review in Appendix \ref{sec:sphere-partition-function} following \cite{Carlip:1994gc}, for the cusp geometry the regularized Einstein-Hilbert action is found to vanish.\footnote{There is an important caveat to this statement: the individual components of the Riemann and Ricci tensors diverge at the cusp singularity $(r=\infty)$. Our computation uses the two derivative gravity action in a regulated cusp geometry with large and small radial cutoffs. Using this regulated action, we see that the contribution from the cusp singularity vanishes. Ideally, the partition function should be computed using worldsheet string theory. However, beyond very special examples the calculation of the sphere partition function in string theory is currently not understood. On the other hand, there are reasons to suspect our computation might be unchanged. Because the geometry is Einstein and three-dimensional, the Riemann tensor can be expressed as a product of two metric tensors. Any higher derivative diffeomorphism invariant constructed from Riemann then turns out to be decaying at the cusp singularity. This suggests that the vanishing seen at two derivatives might persist when higher derivative terms are taken into account in pure gravity.} On the other hand, for thermal AdS$_3$ one finds  
\begin{equation}
e^{Z_{\text{sphere}}}=e^{-I_{\text{on-shell}}}=e^{\frac{\mathtt c_0 \pi t_2}{6}}=|x|^{\frac{\mathtt c_0}{12}}\,, 
\end{equation}
which agrees with the expectation that the dual CFT ground-state energy is $-\frac{\mathtt c_0}{12}$. 

As already mentioned, the argument above is based on the idea that string theory should reproduce semiclassical gravity for small $\alpha'$. While we do not know a first principle computation of the sphere partition function for the very `stringy' $k=1$ setup, one can proceed as in \cite{Eberhardt:2020bgq} and assume that the semi-classical result can be trusted all the way to $k=1$. We are then led to conjecture that also at $k=1$ the cusp geometry sphere partition function vanishes. Notice that this assumption explains why for $\lambda =0$ the one loop contribution to the cusp geometry string partition function \emph{alone} --- with no need to add any further contribution --- could reproduce the full string theory partition function of thermal AdS$_3$ and confirm the claim of \cite{Eberhardt:2020bgq, Eberhardt:2021jvj} that at $k=1$ the string partition function does not depend on the specific choice of bulk geometry. 

Let us then discuss the sphere partition function for the string theory introduced in \eqref{ads3/cft2-deformed} and extend the discussion above to non-vanishing values of $\lambda$. While the computation of the sphere partition function at large $k$ and $\lambda \neq 0$ by holographic renormalization techniques is beyond the scope of this paper, it is reasonable to expect that the deformed sphere  partition can be simply obtained by $\TT$ deforming the $\lambda = 0$ result by the square root formula \eqref{CFTenergies}. This suggests that also at $\lambda \neq 0$ the cusp geometry sphere partition function vanishes. In the following, we will assume that this is indeed the case, at least for $k=1$. We are going to show in Section~\ref{sec:comparison} that this assumption leads to an exact match with the boundary field theory partition function. 

\subsection{Different instanton sectors}

Let us now be more precise about the set of integers $a,b,c,d$ over which the sum in eq.~\eqref{eq:string-one-loop-final} should run. Notice that modular transformations both on the worldsheet and in spacetime only relate terms with the same determinant\footnote{The determinant \eqref{w=det} is identified with the asymptotic winding of the string. In the following, we will thus frequently refer to the determinant \eqref{w=det} as `winding'.}
\begin{equation}
    w = \det \begin{pmatrix} a & b \\ c & d \end{pmatrix} = ad - bc \,.  
    \label{w=det}
\end{equation}
Choosing the values of $w$ entering the  sum in eq.~\eqref{eq:string-one-loop-final} amounts to picking a specific definition of the path integral and to choosing which instanton sectors contribute to the bulk gravity theory. Any collection $\{w_i\} \subset \mathbb Z$ of integer windings would in principle produce a modular invariant partition function. However, while we do not have a good understanding of the worldsheet fusion rules of the $J^+ \bar J^+$ deformed worldsheet theory, we do know that closure of the fusion rules for the $\lambda = 0$ undeformed tensionless string requires to include in the spectrum all the positive integer values of $w \in \mathbb N$ \cite{Eberhardt:2018ouy}. As a consequence, we believe that at least all the positive integer windings $w \in \mathbb N$ should be included  also in the $J^+ \bar J^+$ deformed string theory partition function. Let us thus consider in particular the following possibilities:
\begin{enumerate}
    \item Restrict the path integral sum over only positive integer values of the winding~${w \in \mathbb N}$;  
    \item Let the path integral sum run over all integer values of the winding $w \in \mathbb Z$.
\end{enumerate}
In the first case the string partition function reads 
\begin{tcolorbox}[left=0pt,right=0pt,top=0pt,bottom=10pt]
\begin{equation}
\hspace{-5pt}Z_{\text{string}}\begin{bmatrix}\phi\\\chi\end{bmatrix}(t, \sigma|\lambda) = \hspace{-3pt} \sum_{w=1}^{\infty} p^w \hspace{-5pt} \sum_{\gamma \in \overline{M}_w} \hspace{-5pt}\frac{t_2}{4\lambda} \int_{\mathcal{F}} \frac{\text d^2 \tau}{\tau_2^2} \exp{\Big( \hspace{-3pt}- \frac{\pi |\gamma \cdot t - \tau|^2}{2 (\gamma \cdot_t \lambda) \tau_2}\Big)} Z^{\mathbb{T}^4} \hspace{-3pt}
 \begin{bmatrix}
	    \gamma \cdot \begin{matrix}
	        \phi \\ \chi
	    \end{matrix}
	\end{bmatrix} \hspace{-3pt}(\tau|0) \,, 
\label{positive-det}
\end{equation}
\end{tcolorbox}
\noindent while the second option gives 
\begin{equation}
\hspace{-5pt}Z_{\text{string}}^{\text{np}}\begin{bmatrix}\phi\\\chi\end{bmatrix}(t, \sigma|\lambda) = \hspace{-3pt} \sum_{w\in \mathbb Z} p^{w} \hspace{-5pt} \sum_{\gamma \in \overline{M}_w} \hspace{-5pt}\frac{t_2}{4\lambda} \int_{\mathcal{F}} \frac{\text d^2 \tau}{\tau_2^2} \exp{\Big( \hspace{-3pt}- \frac{\pi |\gamma \cdot t - \tau|^2}{2(\gamma \cdot_t \lambda) \tau_2}\Big)} Z^{\mathbb{T}^4} \hspace{-3pt}
 \begin{bmatrix}
	    \gamma \cdot \begin{matrix}
	        \phi \\ \chi
	    \end{matrix}
	\end{bmatrix} \hspace{-3pt}(\tau|0) \,.
\label{all-det}
\end{equation}
\noindent Let us remind the reader that for the $\lambda = 0$ tensionless string, there is no distinction between the two options above: as discussed around eq.~\eqref{delta-function-identity}, the exponential in eq.~\eqref{string-partition-function} localizes to the Dirac delta 
\begin{equation}
    \delta^2\Bigl( (at+b)-(ct+d)\tau \Bigr) \,, 
    \label{delta-function-lambda=0-2}
\end{equation}
so that for $t_2 >0$ and $\tau_2 > 0$ negative values of $w$ decouple and do not contribute.\footnote{Also contributions from vanishing $w = ad-bc$ decouple, see \cite{Eberhardt:2020bgq}.}

\paragraph{The negative winding sectors} It is intriguing to investigate the physics of the partition function \eqref{all-det} and to understand the nature of the states being counted. In particular, which states are being counted by the negative winding sectors in \eqref{all-det}? To explain this, let us consider the contribution of the $w=-1$ sector, which can be rewritten in the form
\begin{equation}
	\text{Tr}\left[ g q^{\mathcal H_\lambda} \tilde g \bar q^{\tilde{ \mathcal H}_\lambda} \right] = \frac{t_2}{2\lambda} \int_{\mathbb{H}} \frac{\text d^2 \tau} {\tau_2^2} \exp \left( -\frac{\pi |t+\tau|^2}{2\lambda \tau_2} \right) Z^{\mathbb T^4}\begin{bmatrix}
		\phi \\ \chi
	\end{bmatrix}(\tau|0) \,.
\end{equation}
Performing the integral over $\tau$, one identifies the seed Hamiltonian 
\begin{equation}
\begin{aligned}	
\mathcal H_\lambda + \tilde{\mathcal H}_\lambda  &= \frac{1}{2\lambda} \Big(1+\sqrt{1+4\lambda (h+\tilde h) +4 \lambda^2 (h-\tilde h)^2} \, \Big) \,, \\[0.2cm]
\mathcal H_\lambda - \tilde{\mathcal H}_\lambda &= -h + \tilde h \,.
\end{aligned}
\end{equation}
Comparing with eq.~\eqref{eq:energy-formula} and noticing the opposite sign in front of the $1$ inside the bracket, we can rewrite it in terms of the usual $\TT$ deformed energy $H_\lambda$ as
\begin{equation}
\begin{aligned}	
    \mathcal H_\lambda + \tilde{\mathcal H}_\lambda &= \frac{1}{\lambda} + H_\lambda + \tilde H_\lambda \,, \\ 
    \mathcal H_\lambda - \tilde{\mathcal H}_\lambda &=  -H_\lambda + {\tilde H}_\lambda \,. 
\end{aligned}
\label{1-lambda-factor}
\end{equation}
We thus understand that states identified by negative values of the determinant $w$ are heavy for small $\lambda$ and decouple from the spectrum in the $\lambda \to 0$ limit. It also becomes clear that the deformed energy \eqref{1-lambda-factor} is non-perturbative in $\lambda$. A similar computation can be easily carried out for generic negative integer values of winding. 

Let us then discuss the physics associated to the negative winding sectors. As we already mentioned, the determinant $w$ corresponds to the asymptotic winding of the string \cite{Eberhardt:2020bgq}. Its sign tells us whether the string is winding clockwise or anticlockwise around the boundary of $\text{AdS}_3$; opposite orientations of the string correspond to opposite charges under the background $B$ field. For the undeformed $\lambda =0$ tensionless string, the only representations entering the spectrum have $\text{SL}(2, \mathbb R)$ spin $j=\frac{1}{2}$ and correspond to worldsheets covering the $\text{AdS}_3$ asymptotic boundary with positive winding. These positive winding strings are located at the boundary because of a flat potential: the $B$ field exactly balances the tension of the string. On the other hand, since strings of opposite winding have the same tension but opposite charge under the $B$ field, for negative winding strings the potential is no longer flat and they do no longer localize at the boundary of $\text{AdS}_3$. The states generated by their excitations are heavy and in fact as we discussed around eqs.~\eqref{delta-function-lambda=0-1} and \eqref{delta-function-lambda=0-2} they decouple from the spectrum when $\lambda =0$. The effect of the deformation by the marginal operator $J^+ \bar J^+$ is to regularize the infinite energy of negative winding strings and results in the $\lambda^{-1}$ factor in eq.~\eqref{1-lambda-factor}. Indeed, a similar picture for the role of negative winding states has been discussed for $k>1$ bosonic strings in \cite{Giveon:2017myj, Chakraborty:2024mls}. 

\section{Comparison of string and boundary field theory partition functions}
\label{sec:comparison}

In this section we compare the boundary field theory $\TT$ deformed partition function derived in Section~\ref{sec:boundaryfieldtheory} to the  string partition function computed in Section~\ref{sec:worldsheet}. Inspired by the analysis on the worldsheet, in Section~\ref{sec:non-perturbative} we propose a non-perturbative completion of the symmetric orbifold $\TT$ deformed partition function and compare it with various results previously appeared in the literature. 

\subsection{Twisted sectors and positive winding}

In Section~\ref{sec:worldsheet} we derived the grand canonical string partition function~\eqref{positive-det} by including in the spectrum only positive integer values of the winding. It is easy to check that eq.~\eqref{positive-det} exactly matches the boundary field theory $\TT$ deformed partition function computed in Section~\ref{sec:boundaryfieldtheory}, see eq.~\eqref{spacetime-final-pf},
\begin{equation}
    Z_{\text{string}}\begin{bmatrix}\phi\\\chi\end{bmatrix}(t, \sigma|\lambda) = \log\mathfrak{Z} \begin{bmatrix} 	    \phi \\ \chi	\end{bmatrix}(t, \sigma|\lambda) \,. 
\end{equation}
This precise match strongly supports the proposal \eqref{ads3/cft2-deformed} formulated in the Introduction and provides a precise check of various claims which appeared in the literature in recent years. 

\subsection[Non-perturbative completion of \texorpdfstring{$\TT$}{TT} and integer winding]{\boldmath Non-perturbative completion of \texorpdfstring{$\TT$}{TT} and integer winding}
\label{sec:non-perturbative}

As we already discussed at the end of Section~\ref{sec:worldsheet}, it is intriguing to also consider the possibility to define the string theory path integral by including all integer windings $w \in \mathbb Z$. In this case, the string partition function is given by eq.~\eqref{all-det}, which rewritten in the form \eqref{eq:string-one-loop-final} reads
\begin{multline}
Z_{\text{string}}^{\text{np}}\begin{bmatrix}\phi\\\chi\end{bmatrix}(t, \sigma|\lambda) = \sum_{a,b,c,d \, \in \, \mathbb Z}p^{ad-bc}\frac{t_2}{4\lambda}\\
\times \, \int_{\mathcal{F}}\frac{\mathrm{d}^2\tau}{\tau_2^2}\exp\left(-\frac{\pi\left|(at+b)-(ct+d)\tau\right|^2}{2\lambda\tau_2}\right)Z^{\mathbb{T}^4}\begin{bmatrix}a\phi + b\chi \\ c\phi + d\chi\end{bmatrix}(\tau)\,,
\label{Z2-to-manipulate}
\end{multline}
where we remind that $p = \exp(2 \pi i \sigma)$. From the boundary field theory perspective, eq.~\eqref{Z2-to-manipulate} is a non-perturbative completion of eq.~\eqref{spacetime-final-pf}, which we derived in Section~\ref{sec:boundaryfieldtheory}. We hence find it natural to introduce in the boundary field theory a new non-perturbative completion of the $\TT$ kernel formula \eqref{spacetime-final-pf}, defined as  
\begin{multline}
\log\mathfrak{Z}^{\text{np}} \begin{bmatrix} \phi \\ \chi \end{bmatrix}\hspace{-3pt} (t, \sigma|\lambda) = \sum_{a,b,c,d \, \in \, \mathbb Z}p^{ad-bc}\frac{t_2}{4\lambda}\\
\times \, \int_{\mathcal{F}}\frac{\mathrm{d}^2\tau}{\tau_2^2}\exp\left(-\frac{\pi\left|(at+b)-(ct+d)\tau\right|^2}{2\lambda\tau_2}\right)Z^{\mathbb{T}^4}\begin{bmatrix}a\phi + b\chi \\ c\phi + d\chi\end{bmatrix}(\tau)\,. 
\label{np-Z-boundary}
\end{multline}
While the matching of bulk and boundary partition functions 
\begin{equation}
   Z_{\text{string}}^{\text{np}} \begin{bmatrix}\phi\\\chi\end{bmatrix}(t,\sigma|\lambda) =  \log\mathfrak{Z}^{\text{np}} \begin{bmatrix} \phi \\ \chi \end{bmatrix}\hspace{-3pt} (t, \sigma|\lambda)
   \label{np-match}
\end{equation}
is of course built in by definition, we are going to show that for a specific value of the chemical potential $\sigma$, namely 
\begin{equation}
     \sigma = \sigma_{\text{np}} \equiv -\rho_1 + i \frac{t_2}{2 \lambda} \,, \qquad \rho_1 \in \mathbb R \,, 
     \label{sigma-np}
\end{equation}
the non-perturbative definition \eqref{np-Z-boundary} is not new and in fact reduces to the $\TT$ non-perturbative completion introduced and discussed in \cite{Benjamin:2023nts}. Before discussing this, in order to introduce the necessary ingredients, it is beneficial to have a closer look at the $w=1$ sector. Precisely as in the case of the undeformed tensionless string of \cite{Eberhardt:2018ouy}, the $w=1$ sector of the bulk theory exactly matches the untwisted sector of the dual symmetric orbifold. 

\subsection*{\boldmath The untwisted sector and \texorpdfstring{$w=1$}{w=1} winding}

We already reviewed in Section~\ref{sec:boundaryfieldtheory} that given a 2D CFT with partition function $Z(\tau|0)$ the partition function of the $\TT$ deformed field theory takes the form \cite{Dubovsky:2018bmo, Cardy:2018sdv, Hashimoto:2019wct}
\begin{equation}\label{TT-kernel-formula-5}
\begin{split}
	Z(t|\lambda) &= \frac{1}{2} \int_{\mathcal F} \frac{\text d^2 \tau}{\tau_2^2} \, \mathcal K_{w=1}(\tau, t|\lambda) \, Z(\tau|0) \\
 &= \frac{t_2}{2\lambda} \sum_{\gamma \in \text{PSL}(2, \mathbb Z)} \int_{\mathcal F} \frac{\text d^2 \tau} {\tau_2^2} \exp \left( -\frac{\pi |\gamma \cdot t-\tau|^2}{2 \, (\gamma \cdot_t \lambda) \, \tau_2} \right) Z(\tau|0) \,, 
\end{split}
\end{equation}
where we introduced the kernel 
\begin{equation}
    \mathcal K_{w=1}(\tau, t|\lambda) =    \frac{t_2}{\lambda} \sum_{\gamma \in \text{PSL}(2, \mathbb Z)} \exp\left(- \frac{\pi |\gamma \cdot t - \tau|^2}{2 \, (\gamma \cdot_t \lambda) \, \tau_2 }  \right) \,. 
    \label{TT-kernel}
\end{equation} 
It was noticed in \cite{Callebaut:2019omt} that the kernel \eqref{TT-kernel} has exactly the form of the $w=1$ sector of the partition sum of two free bosons
\begin{equation}
X^1, X^2 : \mathbb T^2_\tau \to \mathbb T^2_t   \,,  
\label{two-free-bosons}
\end{equation}
from a worldsheet torus to a spacetime torus with spacetime metric $G_{ab}$ and $B$ field $B_{ab}$ chosen as 
\begin{equation}
    G_{ab} = \frac{1}{2 \lambda} \begin{pmatrix} 1 & t_1 \\ t_1 & t \bar t \end{pmatrix} \,, \qquad B_{ab} = \epsilon_{ab} B =  i \begin{pmatrix} 0 & \frac{t_2}{2 \lambda} \\ -\frac{t_2}{2 \lambda} & 0 \end{pmatrix} \,. 
    \label{G-and-B-im}
\end{equation}
Indeed, for generic values of the metric and $B$ field, the action for the free bosons \eqref{two-free-bosons} reads
\begin{equation}
    S = \frac{1}{2 \pi} \int \text d^2 z \, (G_{ab} + B_{ab}) \partial X^a \bar \partial X^b \,,
    \label{c=2-action}
\end{equation}
and the associated partition function is given by 
\begin{equation}
    \frac{\mathcal K_{c=2}(\rho, t, \tau)}{\tau_2 |\eta(\tau)|^4} \,, 
    \label{c=2-partition-function-with-etas}
\end{equation}
where 
\begin{multline}
    \mathcal K_{c=2}(\rho, t, \tau) = \frac{(\rho - \bar \rho)}{2i} \\
    \times \ \sum_{a,b,c,d \in \mathbb Z} \exp\left( \frac{i \pi}{2 \tau_2 t_2} \left( \rho |at + b-(ct+d) \tau|^2  - \bar \rho |at + b-(ct+d) \bar \tau|^2 \right) \right) \,. 
    \label{c=2-part-funct}
\end{multline}
The complex structure $t$ and the complexified Kahler structure $\rho$ entering eq.~\eqref{c=2-part-funct} can be written in terms of the spacetime metric and $B$ field as\footnote{While we will always assume a real metric and therefore $\bar t$ is the complex conjugate of $t$, in the discussion below it proves useful to also let the $B$ field take complex values. As a consequence, generically $\bar \rho$ is not the complex conjugate of $\rho$, but it is instead defined as in \eqref{t-and-rho-GB}.}
\begin{equation}
t =  \frac{G_{12}}{G_{11}} + i \frac{\sqrt{\det G}}{G_{11}} \,, \qquad \rho = B + i \sqrt{\det G} \,, \qquad \bar \rho = B - i \sqrt{\det G}  \,, 
\label{t-and-rho-GB}
\end{equation}
and the free bosons obey the twisted boundary conditions
\begin{align}
X^1(z+\tau) &= X^1(z) + b \,, & \qquad X^2(z+\tau) &= X^2(z) + a \,, \\
X^1(z+1) &= X^1(z) + d \,, & \qquad X^2(z+1) &= X^2(z) + c \,. 
\end{align}
Indeed, it is not difficult to check that restricting to the $w=ad-bc=1$ sector, and making use of eqs.~\eqref{G-and-B-im} and \eqref{t-and-rho-GB}, eq.~\eqref{c=2-part-funct} reduces to the kernel~\eqref{TT-kernel} \cite{Callebaut:2019omt}. 

Specializing to bosonic 2D CFTs with central charge $c=24$, in \cite{Callebaut:2019omt} eq.~\eqref{TT-kernel-formula-5} was interpreted from a string worldsheet perspective. The authors suggested that any $\TT$ bosonic deformed field theory with $c=24$ is dual to a non-critical string theory. The worldsheet CFT defining the latter is given by the $w=1$ sector of two compact free bosons with spacetime metric and $B$ field as in \eqref{G-and-B-im}, together with the undeformed $c=24$ CFT itself and the usual $c=-26$ $bc$ ghosts of bosonic string theory. Indeed, the Dedekind eta function factors entering the free boson partition function \eqref{c=2-partition-function-with-etas} cancel against the contribution of the $bc$ ghosts and the string partition function of such bosonic string theory reproduces eq.~\eqref{TT-kernel-formula-5}. 

While in this manuscript we only consider superstrings and $\TT$-deformed supersymmetric CFTs with central charge $c=6$, let us explain how a mechanism similar to the one advocated in \cite{Callebaut:2019omt} indeed emerges from the worldsheet of the $J^+ \bar J^+$ deformed tensionless string, when considering the winding one sector $w = ad-bc =1$. In fact, writing 
\begin{equation}
    \gamma = \frac{X^1 + t X^2}{\sqrt 2}  \,, \qquad  \bar \gamma = \frac{X^1 + \bar t X^2}{\sqrt 2} \,,
\end{equation}
the action \eqref{S-gamma-bar-gamma} takes exactly the form \eqref{c=2-action} with spacetime metric and $B$ field as in \eqref{G-and-B-im}. Isolating the $w=1$ contribution of the string partition function we computed in Section~\ref{sec:worldsheet}, see eq.~\eqref{string-partition-function-w=1}, we exactly recover the untwisted sector of the boundary field theory, see eq.~\eqref{eq:w-step-2}. This precisely mimics the tensionless string duality of \cite{Eberhardt:2018ouy}: the $w=1$ sector of the string exactly matches the untwisted sector of the dual symmetric orbifold. 

\subsection*{\boldmath Comparison with the $\TT$ non-perturbative completion of \cite{Benjamin:2023nts}}

Various non-perturbative completions of partition function for  $\TT$ deformed theories have been discussed in the literature~\cite{Jiang:2019epa,Griguolo:2022xcj, Benjamin:2023nts}. Let us briefly review the one we will be mainly interested in, explored by Benjamin, Collier, Kruthoff, Verlinde and Zhang in \cite{Benjamin:2023nts}. There the authors propose a natural extension of the DMVV formula for $\TT$ deformed symmetric product CFTs, by defining the free energy 
\begin{equation}
    F(\rho_{\text{np}}, t) = \frac{1}{2} \int_{\mathcal{F}} \frac{\text d^2 \tau}{\tau_2^2} \mathcal K_{c=2}(\rho_{\text{np}}, t, \tau) Z^{\text{seed}} (\tau|0) \,,
    \label{free-energy-BCKVZ}
\end{equation}
where $\rho_{\text{np}}$ reads
\begin{equation}
\rho_{\text{np}} \equiv  \rho_1 + i \frac{t_2}{2 \lambda} \,, \qquad \rho_1 \in \mathbb R \,.
\label{rho-np}
\end{equation}
Contact with the usual DMVV formula~\cite{Dijkgraaf:1996xw, Maldacena:1999bp} can be made by noticing that the kernel \eqref{c=2-part-funct} entering equation \eqref{free-energy-BCKVZ} can be expressed as a sum over terms with torus wrapping $N$~\cite{Benjamin:2021ygh},
\begin{equation}
    \mathcal K_{c=2}(\rho, t, \tau) = \sum_{N>0} \hat T_N \mathcal K_{+1} + \mathcal K_0 + \sum_{N>0} \hat T_N  \mathcal K_{-1} \,, 
\end{equation}
where\footnote{Since this will not play a role in our analysis, we refer the reader to~\cite{Benjamin:2021ygh} and \cite{Benjamin:2023nts} for the precise definition of $\mathcal K_0$ and an analysis of its features.}
\begin{align}
    \mathcal K_0(\rho, t, \tau) &= \rho_2 + 2 \rho_2 \sum_{n=1}^\infty \sum_{\gamma, \tilde \gamma \in \Gamma_\infty \setminus \text{PSL}(2, \mathbb Z)} e^{-\frac{\pi n^2 \rho_2}{(\gamma \cdot t)_2 (\tilde \gamma \cdot \tau)_2}} \,, \\
    \mathcal K_{\pm 1}(\rho, t, \tau) &= 2\rho_2 \sum_{\gamma \in \text{PSL}(2,\mathbb Z)} e^{\frac{i \pi}{2} \left( \frac{\rho}{t_2 (\gamma \cdot \tau)_2} |t\mp \gamma \cdot \tau|^2 - \frac{\bar \rho}{t_2 (\gamma \cdot \tau)_2} |t\mp \gamma \cdot \bar \tau|^2  \right)} \,, 
\end{align}
and $\hat T_N$ denotes the Hecke operator acting on a weight-0 modular form $\phi(\rho, t)$ as 
\begin{equation}
    \hat T_N\phi(\rho, t) = \frac{1}{N} \sum_{\substack{ad=N, \, d>0 \\ b \, \text{mod} \, d }} \phi \left(N \rho, \frac{a t+b}{d} \right) \,. 
\end{equation}

The partition sum \eqref{c=2-part-funct} satisfies a number of remarkable properties, including invariance under modular transformations of $t$, invariance under modular transformations of $\rho$ and a triality symmetry under permutations of $\tau, t$ and $\rho$ \cite{Dijkgraaf:1987jt}. The free energy \eqref{free-energy-BCKVZ} inherits many of the nice properties of the partition sum \eqref{c=2-part-funct}: it is invariant under modular transformations of $\rho$, under modular transformations of $t$, under the exchange of $t$ and $\rho$ and various additional properties for which we refer to \cite{Benjamin:2023nts}. The property most relevant for us, is the invariance under the map $\lambda \to \frac{1}{4 \lambda}$. Thanks to this symmetry, positive values of $\lambda$ that would render imaginary the square root in eq.~\eqref{double-trace}, can be interpreted as the mirror image of positive values of $\lambda$ for which the square root remains real. 

Let us now specialize the non-perturbative completion \eqref{free-energy-BCKVZ} by choosing as seed theory the $\mathbb T^4$ CFT. By including dependence on the spin structure, we obtain
\begin{multline}
F\begin{bmatrix} \phi \\ \chi \end{bmatrix}\hspace{-3pt}(\rho_{\text{np}}, t) = \frac{t_2}{4 \lambda} \sum_{a,b,c,d \in \mathbb Z} \int_{\mathcal{F}} \frac{\text d^2 \tau}{\tau_2^2} \, Z^{\mathbb{T}^4} \hspace{-3pt} \begin{bmatrix} \begin{matrix} a \phi + b \chi \\ c \phi + d \chi \end{matrix} \end{bmatrix} \hspace{-3pt}(\tau|0)  \\
\times \, \exp\left( \frac{i \pi}{2 \tau_2 t_2} \left( \rho_\text{np} |at + b-(ct+d) \tau|^2  - \bar \rho_\text{np} |at + b-(ct+d) \bar \tau|^2 \right) \right)  \,. 
\label{np-completion-spin-structure}
\end{multline}
Some simple algebra shows that for a specific choice of the chemical potential $\sigma$, namely 
\begin{equation}
     \sigma = \sigma_{\text{np}} \equiv -\rho_1 + i \frac{t_2}{2 \lambda} \,,
     \label{sigma-np-2}
\end{equation}
the $\TT$ non-perturbative completion that we proposed in \eqref{np-Z-boundary} reduces to the one studied in \cite{Benjamin:2023nts}, i.e.
\begin{equation}
    \log\mathfrak{Z}^{\text{np}} \begin{bmatrix} \phi \\ \chi \end{bmatrix}\hspace{-3pt} (t, \sigma_{\text{np}}|\lambda) = F\begin{bmatrix} \phi \\ \chi \end{bmatrix}\hspace{-3pt}(\rho_{\text{np}}, t) \,. 
\end{equation}
Let us also briefly comment on what is special about choosing the chemical potential $\sigma$ as in eq.~\eqref{sigma-np-2}. The effect of the $B$ field introduced around eq.~\eqref{eq:topological-term} to define the grand-canonical partition function, is to replace the two free bosons action action \eqref{S-gamma-bar-gamma} with 
\begin{equation}
    S_\lambda^{\gamma \bar \gamma} \mapsto \frac{1}{2 \pi} \int \text d^2 z \left(  G_{ab} \partial X^a \bar \partial X^b + \epsilon_{ab}(B - \sigma) \partial X^a \bar \partial X^b \right) \,,
    \label{S-total-B-field}
\end{equation}
where
\begin{equation}
    B = \frac{i \, t_2}{2 \lambda} \,, 
\end{equation}
is given by eq.~\eqref{G-and-B-im}. We thus see in which sense the value \eqref{sigma-np-2} of the grand-canonical chemical potential $\sigma$ is special: it is the only value for which the total $B$ field entering the two free bosons action \eqref{S-total-B-field} is real. 

\vskip 0.4 in

\section*{Acknowledgements}

We thank Luis Apolo, Nathan Benjamin, Davide Bufalini, Soumangsu Chakraborty, Scott Collier, Lorenz Eberhardt, Matthias Gaberdiel, Shota Komatsu, Nicolas Kovensky, Per Kraus, David Kutasov, Emil Martinec, Ruben Monten, Greg Moore, Beat Nairz, Alessandro Sfondrini, Wei Song, Vit Sriprachyakul and Houri Christina Tarazi for useful discussions. We are especially grateful to Lorenz Eberhardt and Matthias Gaberdiel for helpful comments on a draft of this paper. AD and BK thank the Institut Pascal, participants and organizers of the workshop `Speakable and unspeakable in quantum gravity' for their hospitality during the final stages of this work and for very stimulating discussions. AD~acknowledges support from the Mafalda \& Reinhard Oehme Fellowship. The work of BK was supported by STFC consolidated grants ST/T000694/1 and ST/X000664/1. The work of KN was supported by the Swiss National Science Foundation through a personal grant and via the NCCR SwissMAP. The work of SS is supported in part by NSF Grant No. PHY2014195.

\appendix

\section{Theta functions}
\label{app:theta}

In this appendix, we list our conventions for the Theta functions. We define Theta functions as\footnote{Our conventions are related to those in \cite{Blumenhagen:2013fgp} by $\vartheta_{\text{here}}\begin{bmatrix} \alpha \\ \beta \end{bmatrix} = \vartheta_{\text{there}}\begin{bmatrix} \alpha + \tfrac{1}{2} \\ \beta + \tfrac{1}{2} \end{bmatrix}$.}
\begin{equation}
    \vartheta \begin{bmatrix} \alpha \\ \beta \end{bmatrix}(u,\tau) = \sum_{n \in \mathbb Z} \exp \Bigl( \pi i (n+\alpha + \tfrac{1}{2})^2 \tau + 2 \pi i(n+\alpha+\tfrac{1}{2})(u+\beta + \tfrac{1}{2})  \Bigr) \,,
    \label{Theta-def}
\end{equation}
for any $\alpha, \beta \in \mathbb R$ and $\text{Im}(\tau) >0$. When $-1 \leq \chi, \phi \leq 0$ we also have \cite{Blumenhagen:2013fgp}
\begin{multline}
    \vartheta \begin{bmatrix} \chi \\ \phi \end{bmatrix}(0,\tau) \equiv \vartheta \begin{bmatrix} \chi \\ \phi \end{bmatrix}(\tau) = e^{2 \pi i (\phi + \frac{1}{2})(\chi + \frac{1}{2})} q^{\frac{1}{2}(\chi+\frac{1}{2})^2} \\
    \times \, \prod_{n=1}^\infty (1-q^n)(1+e^{2 \pi i (\phi + \frac{1}{2})} q^{n+\chi})(1+e^{-2\pi i (\phi + \frac{1}{2})}q^{n - 1 - \chi}) \,.
    \label{theta-theta-phi}
\end{multline}
For $\alpha, \beta \in \{ 0, -\tfrac{1}{2}\}$, the Theta functions \eqref{Theta-def} have special names,
\begin{align}
    \vartheta \begin{bmatrix} 0 \\ 0 \end{bmatrix}(u,\tau) &= \vartheta_1(u,\tau) = i e^{\pi i u} q^{\frac{1}{8}} \prod_{n=1}^\infty (1-q^n)(1- e^{2 \pi i u} q^n) (1- e^{-2 \pi i u} q^{n-1}) \,, \label{theta1} \\
    \vartheta \begin{bmatrix} 0 \\ -\tfrac{1}{2} \end{bmatrix}(u,\tau) &= \vartheta_2(u,\tau) = e^{\pi i u} q^{\frac{1}{8}} \prod_{n=1}^\infty (1-q^n)(1+ e^{2 \pi i u} q^n) (1+ e^{-2 \pi i u} q^{n-1}) \,, \\
     \vartheta \begin{bmatrix} -\tfrac{1}{2} \\ -\tfrac{1}{2} \end{bmatrix}(u,\tau) &= \vartheta_3(u,\tau) = \prod_{n=1}^\infty (1-q^n)(1 + e^{2 \pi i u} q^{n-\frac{1}{2}}) (1+ e^{-2 \pi i u} q^{n-\frac{1}{2}}) \,, \\
    \vartheta \begin{bmatrix} -\tfrac{1}{2} \\ 0 \end{bmatrix}(u,\tau) &= \vartheta_4(u,\tau) = \prod_{n=1}^\infty (1-q^n)(1- e^{2 \pi i u} q^{n-\frac{1}{2}}) (1- e^{-2 \pi i u} q^{n-\frac{1}{2}}) \,. \label{theta4}
\end{align}
For $\alpha, \beta \in \{ 0, -\tfrac{1}{2}\}$, we can summarize eqs.~\eqref{theta1}-\eqref{theta4} as
\begin{multline} \label{theta-abz}
    \vartheta \begin{bmatrix} \beta \\ \alpha \end{bmatrix}(u,\tau) = e^{2 \pi i (\frac{1}{2}+\beta)(\alpha+\frac{1}{2}+u)} q^{\frac{1}{2}(\beta + \frac{1}{2})^2} \\ 
    \times \, \prod_{n=1}^\infty (1-q^n)(1+e^{2 \pi i (\alpha + \frac{1}{2} + u)} q^{n+\beta})(1+e^{-2\pi i (\alpha + \frac{1}{2}+u)}q^{n - 1 - \beta}) \,.
\end{multline}
Defining $\chi$ and $\phi$ as
\begin{equation} \label{eq:u-in-terms-phi-theta}
    u = (\chi-\beta) \tau -(\phi - \alpha) \,,
\end{equation}
and comparing eqs.~\eqref{theta-theta-phi} and \eqref{theta-abz}, it is easy to check that
\begin{equation}
    \left| \vartheta \begin{bmatrix} \chi \\ -\phi \end{bmatrix}(0,\tau) \right|^2 =  \left| \vartheta \begin{bmatrix} \beta \\ -\alpha \end{bmatrix}(u,\tau) \right|^2 e^{-2 \pi \frac{u_2^2}{\tau_2}} \,,
    \label{theta-phi-trick}
\end{equation}
or equivalently
\begin{equation}
\left| \vartheta \begin{bmatrix} \beta \\ -\alpha \end{bmatrix}(u,\tau) \right|^2 = \left| \vartheta \begin{bmatrix} \chi \\ -\phi \end{bmatrix}(0,\tau) \right|^2 e^{2 \pi (\chi-\beta )^2 \tau_2} \,. \\[0.4cm]
\end{equation}

\section{Modular properties}
\label{app:Modular properties}

In this appendix, we collect various modular properties that we have used throughout the text. We begin with the modular transformation of the torus modular parameter $t$, and the real spin structures,
\begin{equation} \label{eq:mod-transformations}
    t\mapsto \gamma\cdot t \equiv \frac{at+b}{ct+d} \,, \qquad \begin{bmatrix} \phi \\ \chi \end{bmatrix} \mapsto \gamma \cdot \begin{bmatrix} \phi \\ \chi \end{bmatrix} \,, \qquad \gamma=\begin{pmatrix}
        a & b \\
        c & d
    \end{pmatrix} \in \text{PSL}(2,\mathbb{Z}) \,,
\end{equation}
where in the second equation $(\gamma\ \cdot \ )$ means matrix multiplication on the column vector.

Let us now discuss the modular behavior of the partition functions defined in eqs.~\eqref{T4-part-fun-pi}. Using the identity \eqref{theta-phi-trick}, it is straightforward to rewrite them as in eqs.~\eqref{z-to-phi-theta}. Therefore, it is sufficient to consider the partition function defined in eq.~\eqref{Z-sp-T4}, which we rewrite here,
\begin{equation}
    Z^{\mathbb T^4} \begin{bmatrix} \phi \\ \chi \end{bmatrix}(t|0) = \frac{\Theta(t)}{\left| \eta(t)  \right|^{12}} \left| \vartheta \begin{bmatrix} \chi \\ -\phi \end{bmatrix}(t) \right|^4 \,.
\end{equation}
The Theta functions obey the identity \cite{Blumenhagen:2013fgp}
\begin{equation}
    \Bigg|\frac{1}{\eta(\gamma\cdot t)} \vartheta \begin{bmatrix} \chi \\ -\phi \end{bmatrix}(\gamma\cdot t)\Bigg|^2 = \Bigg|\frac{1}{\eta(t)} \vartheta \begin{bmatrix} -c\phi+a\chi \\ -d\phi+b\chi \end{bmatrix}(t)\Bigg|^2 \,.
    \label{theta-modular-trsf}
\end{equation}
Using this, one can show that
\begin{equation}
	Z^{\mathbb T^4}  \begin{bmatrix} \gamma \cdot \begin{matrix} \phi \\ \chi \end{matrix} \end{bmatrix} ( \gamma \cdot t|0)
 = Z^{\mathbb T^4} \begin{bmatrix} \phi \\ \chi \end{bmatrix}(t|0) \,,\quad \gamma \in \text{PSL}(2,\mathbb{Z}) \,. \label{eq:z-modular-property}
\end{equation}
In passing let us mention the modular property of the exponential $e^{- \frac{\pi}{t_2}u_2^2}$ in eqs.~\eqref{T4-part-fun-pi},
\begin{equation}
    \gamma: \ \frac{\pi u_2^2}{t_2} \ \mapsto \  \frac{\pi u_2^2}{t_2} + \frac{\pi i c u^2}{2(c t + d)} - \frac{\pi i c \bar u^2}{2(c \bar t + d)} \,.
\end{equation}
Using this, it is easy to derive the familiar modular transformation of partition functions defined without the insertion of the exponential factor $e^{- \frac{\pi}{t_2}u_2^2}$, for example,
\begin{equation}
	Z^{\mathbb T^4}_{\widetilde{\text R}} (\gamma\cdot u,\gamma \cdot t)
 = \Big|e^{\frac{\pi i c u^2}{2(c t + d)}}\Big|^2 Z^{\mathbb T^4}_{\widetilde{\text R}} (u,t) \,,\quad \gamma \in \text{PSL}(2,\mathbb{Z}) \,, \label{eq:z-modular-property-chemical-potential}
\end{equation}
where $u$ denotes the chemical potential.

Finally, we consider the chemical potential defined in eq.~\eqref{eq:u-in-terms-phi-theta}. By a direct calculation and using the transformation in eq.~\eqref{eq:mod-transformations}, one can show that
\begin{equation}
	\gamma: \ u\mapsto \frac{u}{ct+d} \,, 
\end{equation}
which is the usual modular transformation of the chemical potential.

\section{Free fermion partition functions}
\label{app:path-integral-free-fermions}

In this appendix we review the path integral derivation of the partition function for complex free fermions. We comment on modular properties and compare the result with the operator formalism. 

Consider a complex free fermion
\begin{equation}
    \psi^+(x_1) \psi^-(x_2) \sim \frac{1}{x_1 - x_2} \ , \qquad (\psi^+)^* = \psi^- \ , 
\end{equation}
together with its anti-holomorphic analogue $\tilde \psi^+$, $\tilde \psi^-$. Let us label points on the torus by
\begin{equation}
    x = \sigma_1 + t \sigma_2 \,, \qquad     \bar x = \sigma_1 + \bar t \sigma_2 \,, 
\end{equation}
where $\sigma_1, \sigma_2 \in [0, 2\pi)$ parametrize the two cycles, and assume periodic boundary conditions 
\begin{equation}
\begin{aligned}
    \psi^\pm(\sigma_1 + 2 \pi, \sigma_2) &= \psi^\pm(\sigma_1, \sigma_2) \,,      &   \qquad     \tilde \psi^\pm(\sigma_1 + 2 \pi, \sigma_2) &= \tilde \psi^\pm(\sigma_1, \sigma_2) \,, \\
\psi^\pm(\sigma_1, \sigma_2 + 2 \pi) &= \psi^\pm(\sigma_1, \sigma_2) \,,  &   \qquad    \tilde \psi^\pm(\sigma_1, \sigma_2 + 2 \pi) &= \tilde \psi^\pm(\sigma_1, \sigma_2) \,. 
\end{aligned}    
\end{equation}
The discussion for the other spin structures is completely analogous. In the operator formalism the partition function of these free fermions reads 
\begin{equation}
    Z^{\text{O.F.}} = \text{Tr}_{\tilde{\text R}}(q^{L_0 - \frac{1}{24}} \, e^{2 \pi i \, u \, j_0} \, \bar q^{\bar L_0 - \frac{1}{24}} \, e^{-2 \pi i \, \bar u \, \bar j_0}) = \frac{1}{\left| \eta(t)  \right|^{2}} \left| \vartheta \begin{bmatrix} 0 \\ 0 \end{bmatrix}(u,t) \right|^2 \,.
    \label{ZOF}
\end{equation}
As we saw in Appendix~\ref{app:Modular properties} the partition function \eqref{ZOF} is not modular invariant. On the other hand, since in the path integral a modular transformation amounts to a change of coordinates followed by a Weyl transformation, we expect a modular invariant result. Let us see how this comes about. The free fermion action reads
\begin{equation}
    S_{\text{free}} = \frac{1}{\pi} \int \text d^2 x \left[ \psi^+ \partial_{\bar x} \psi^- + \tilde \psi^+\partial_x \tilde \psi^- \right] \,.
\end{equation}
Introducing a constant background gauge field, it can be promoted to
\begin{equation}
    S_2 = \frac{1}{\pi} \int \text d^2 x \left[ \psi^+(\partial_{\bar x} - A_{\bar x}) \psi^- + \tilde \psi^+(\partial_x + A_x) \tilde \psi^- \right] \,. 
    \label{S2}
\end{equation}
We are interested in computing the associated path integral
\begin{equation}
    Z_2^{\text{P.I.}} = \int \mathcal D \psi^+ \mathcal D \psi^- \mathcal D \tilde \psi^+ \mathcal D \tilde \psi^+ e^{-S_2} \,. 
    \label{Z1PI}
\end{equation}
The background gauge fields $A_{\bar x}$ and $A_x$ entering eq.~\eqref{S2} are related to the chemical potentials $u$ and $\bar u$ as 
\begin{equation}
    A_{\bar x} = -\frac{u}{2 \tau_2} \,, \qquad      A_x = -\frac{\bar u}{2 \tau_2} \,. 
\end{equation}
As we are going to see momentarily, it is with this choice of normalization that the linear terms in $u$ and $\bar u$ in \eqref{ZOF} and \eqref{Z1PI} agree. In the following it proves useful to introduce the notation
\begin{equation}
    u = \chi t - \phi \,, 
\end{equation}
and the real function
\begin{equation}
    \Phi[u](x) = \frac{i}{2\tau_2}(\bar x u - x \bar u) = -\sigma_1 \chi - \sigma_2 \phi \,, 
\end{equation}
in terms of which the background gauge fields read 
\begin{equation}
    A_{\bar x} = i \partial_{\bar x} \Phi \,, \qquad      A_x = -i \partial_x \Phi \,.  
    \label{Ax}
\end{equation}
One can then rewrite the action \eqref{S2} as
\begin{equation}
    S_2 = \frac{1}{\pi} \int \text d^2 x \left[ \eta^+ \partial_{\bar x}\eta^- + \tilde \eta^+ \partial_x \tilde \eta^- \right] \,,
    \label{S2-eta}
\end{equation}
where
\begin{equation}
\eta^{\pm} = e^{\pm i \Phi} \psi^{\pm} \,, \qquad \tilde \eta^{\pm} = e^{\pm i \Phi} \tilde \psi^{\pm} \,. 
\end{equation}
Notice that while the fermions $\psi^\pm$ and $\tilde \psi^\pm$ are periodic, $\eta^\pm$ and $\tilde \eta^\pm$ obey the twisted boundary conditions
\begin{subequations}
\begin{align}
\eta^\pm(\sigma_1 + 2\pi, \sigma_2) & = e^{\mp 2 \pi i \chi} \, \eta^\pm(\sigma_1, \sigma_2)  \,, & \quad \tilde \eta^\pm(\sigma_1 + 2\pi, \sigma_2) & = e^{\mp 2 \pi i \chi} \, \tilde \eta^\pm(\sigma_1, \sigma_2)  \,, \label{first-bc-eta} \\
\eta^\pm(\sigma_1, \sigma_2 + 2\pi) & = e^{\mp 2 \pi i \phi} \, \eta^\pm(\sigma_1, \sigma_2)  \,, & \quad \tilde \eta^\pm(\sigma_1, \sigma_2 + 2\pi) & = e^{\mp 2 \pi i \phi} \, \tilde \eta^\pm(\sigma_1, \sigma_2)  \,. \label{second-bc-eta}
\end{align}
\end{subequations}
The partition function for twisted fermions has been computed in \cite{Alvarez-Gaume:1986rcs}, see also~\cite{Blumenhagen:2013fgp} for a review. It reads
\begin{align}
 Z_2^{\text{P.I.}} = \int \mathcal D \eta^+ \mathcal D \eta^- \mathcal D \tilde \eta^+ \mathcal D \tilde \eta^+ e^{-S_2} = \text{Tr}\left[q^{h} g \, \bar q^{\tilde h} \tilde g \right] = \frac{1}{\left| \eta(t)  \right|^{2}} \left| \vartheta \begin{bmatrix} \chi \\ -\phi \end{bmatrix}(t) \right|^2 \,, 
 \label{eta-pi}
\end{align}
where $h$ is the holomorphic Hamiltonian
\begin{equation}
    h = L_0 -\frac{1}{24} = \sum_{m=1}^\infty (m+\chi) : b_{m+\chi}^\dagger b_{m+\chi}: + \frac{1}{2} \left(\chi-\tfrac{1}{2}\right)^2 - \frac{1}{24} \,. 
    \label{h}
\end{equation}
Similarly for the anti-holomorphic sector. In \eqref{eta-pi}, $g$ and $\tilde g$ are twist operators implementing the boundary condition \eqref{second-bc-eta}. The boundary condition \eqref{first-bc-eta} instead determines a quantization of the free fermion modes $b_{m+\chi}$ and $b_{m+\chi}^\dagger$ in units of $m + \chi \in \mathbb Z + \chi$. In fact, the zero point energy $\frac{1}{2}(\chi -\frac{1}{2})^2$ in \eqref{h} arises as a normal ordering constant of the twisted free fermions, see \cite{Blumenhagen:2013fgp}. Finally, using the identity \eqref{theta-phi-trick}, one obtains
\begin{equation}
    Z_2^{\text{P.I.}} = \int \mathcal D \eta^+ \mathcal D \eta^- \mathcal D \tilde \eta^+ \mathcal D \tilde \eta^+ e^{-S_2} =  \frac{e^{-\frac{2 \pi}{\tau_2}u_2^2}}{\left| \eta(t)  \right|^{2}} \left| \vartheta \begin{bmatrix} 0 \\ 0 \end{bmatrix}(u,t) \right|^2 \,. 
    \label{Z2}
\end{equation}
We see that the result of the path integral computation differs from eq.~\eqref{ZOF} for a factor $e^{-\frac{2 \pi}{\tau_2}u_2^2}$. It is easy to check that thanks to this factor the partition function \eqref{Z2} is modular invariant, see Appendix~\ref{app:Modular properties}.

Alternatively to eq.~\eqref{S2}, one can couple the background gauge field as
\begin{equation}
    S_1 = \frac{1}{\pi} \int \text d^2 x \left[ \psi^+(\partial_{\bar x} + A_{\bar x}) \psi^- + \tilde \psi^+(\partial_x + A_x) \tilde \psi^- \right] \,. 
    \label{S1}
\end{equation}
Also in this case the action can be brought into the form 
\begin{equation}
    S_1 = \frac{1}{\pi} \int \text d^2 x \left[ \eta^+ \partial_{\bar x}\eta^- + \tilde \eta^+ \partial_x \tilde \eta^- \right] \,,
    \label{S1-eta}
\end{equation}
by defining the twisted fermions as 
\begin{equation}
\eta^{\pm} = e^{\mp i \Phi} \psi^{\pm} \,, \qquad \tilde \eta^{\pm} = e^{\pm i \Phi} \psi^{\pm} \,. 
\label{eta-def-1}
\end{equation}
This time the `rotation' \eqref{eta-def-1} is anomalous and the path-integral measure is not invariant: an axial anomaly arises, 
\begin{equation}
    \mathcal D \psi^+ \mathcal D \psi^- \mathcal D \tilde \psi^+ \mathcal D \tilde \psi^+ = \mathcal D \eta^+ \mathcal D \eta^- \mathcal D \tilde \eta^+ \mathcal D \tilde \eta^+ \exp \left( \frac{2\pi}{\tau_2}|u|^2 \right) \,.  
    \label{axial anomaly}
\end{equation}
The computation of the path integral partition function then proceeds as above and taking into account the anomaly one finds
\begin{equation}
    Z_1^{\text{P.I.}} = \int \mathcal D \eta^+ \mathcal D \eta^- \mathcal D \tilde \eta^+ \mathcal D \tilde \eta^+ e^{-S_1}  =\frac{e^{\frac{2 \pi}{\tau_2}u_1^2}}{\left| \eta(t)  \right|^{2}} \left| \vartheta \begin{bmatrix} 0 \\ 0 \end{bmatrix}(u,t) \right|^2 \,. 
    \label{Z1}
\end{equation}
The path integral this time differs from the operator formalism partition function \eqref{ZOF} by a factor $e^{\frac{2 \pi}{\tau_2}u_1^2}$. Also in this case, the path integral partition function \eqref{Z1} is modular invariant. 

To sum up, the path integral for a complex free fermion coupled to a constant background gauge field differs from the partition function computed in the operator formalism by a factor $e^{-\frac{2 \pi}{\tau_2}u_2^2}$ or $e^{\frac{2 \pi}{\tau_2}u_1^2}$, depending on the definition of the path integral one adopts. 

\section{The sphere partition function}
\label{sec:sphere-partition-function}

In this appendix following \cite{Carlip:1994gc} we review the computation of the on-shell Einstein-Hilbert action of the cusp geometry and thermal AdS$_3$ geometries. 

\subsection*{The cusp geometry}

\begin{figure}
\centering
\begin{tikzpicture}[scale = 0.75]
\fill[gray, opacity = 0.2] (-0.5,1.25+0.3125) to[out = -17, in = 179] (8,-0.15) to[out = 187, in = 30] (2.5,-2.25-0.3125) -- (-0.5,-1.75+0.3125);
\draw[thick] (-1,-2.5+0.5) -- (-1,3.5-0.75-0.15) -- (3,2-0.75-0.15) -- (3,-4+0.5) -- (-1,-2.5+0.5);
\draw[thick] (-0.5,-1.75+0.3125) -- (-0.5,1.25+0.3125) -- (2.5,0.75-0.3125) -- (2.5,-2.25-0.3125) -- (-0.5,-1.75+0.3125);
\draw[thick] (-0.5,-1.75+0.3125) to[out = 10, in = 186] (8,-0.15);
\draw[thick] (2.5,-2.25-0.3125) to[out = 30, in = 189] (8,-0.15);
\draw[thick] (-0.5,1.25+0.3125) to[out = -17, in = 179] (8,-0.15);
\draw[thick] (2.5,0.75-0.3125) to[out = -15, in = 182] (8,-0.15);
\node[above, rotate = -20.55] at (1,2) {\small $r=0$};
\node[right] at (8,-0.15) {\small $r=\infty$};
\end{tikzpicture}
\caption{The fundamental domain of the cusp geometry. The conformal boundary at $r=0$ is a torus formed by identifying $\gamma\sim\gamma+1\sim\gamma+t$. There is a `cusp' of zero volume at $r=\infty$.}
\label{fig:cusp-geometry}
\end{figure}
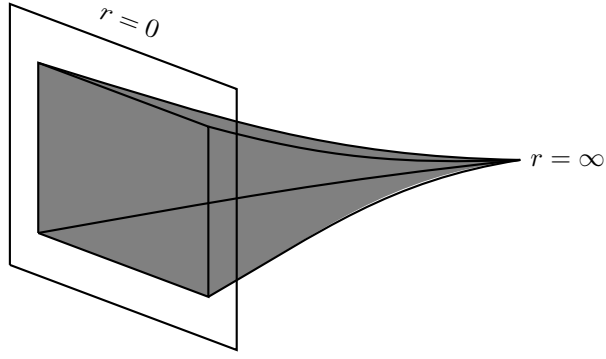

For the cusp geometry, the regularized Einstein-Hilbert action is found to vanish. This can be seen as follows. The cusp geometry is defined by the coordinate system
\begin{equation}
\mathrm{d}s^2=\frac{\mathrm{d}r^2+\mathrm{d}\gamma\,\mathrm{d}\bar\gamma}{r^2}\,,
\end{equation}
in units with $L_{\text{AdS}}=1$. Here, $\gamma$ takes values in $\mathbb{T}^2_t$ and $r\in (0,\infty)$. The `cusp' is located at $r\to\infty$, while the boundary is located at $r\to 0$, see Figure \ref{fig:cusp-geometry}.

In order to calculate the EH action on the cusp geometry, we introduce two cutoffs, one at $r=\varepsilon$ near the boundary and one at $r=\Lambda$ near the cusp. The regularized EH action is
\begin{equation}
I(\varepsilon,\Lambda)=-\frac{1}{16\pi G}\int_M\mathrm{d}^3x\sqrt{g}(R+2)+\frac{1}{8\pi G}\int_{\partial M}\mathrm{d}^2x\sqrt{h}K+\frac{1}{8\pi G}\int_{\partial M}\mathrm{d}^2x\sqrt{h}\,,
\end{equation}
where $M$ is the cusp geometry cut off at $\varepsilon<r<\Lambda$, the boundary $\partial M$ is the two tori at $r=\varepsilon$ and $r=\Lambda$, and the last term is a counterterm added to cancel the infinite-volume divergence. 

The extrinsic curvature can be calculated, and the result is
\begin{equation}
\begin{split}
K=-2\,,&\qquad \text{at} \qquad  r=\varepsilon\,,\\
K=2\,,&\qquad \text{at} \qquad r=\Lambda\,.
\end{split}
\end{equation}
Thus, the regularized Einstein Hilbert action can be computed and we find
\begin{equation}
\begin{split}
I(\varepsilon,\Lambda)&=-\frac{1}{16\pi G}\int_{M}\sqrt{g}(R+2)+\frac{1}{8\pi G}\int_{\partial M}\sqrt{h}K+\frac{1}{8\pi G}\int_{\partial M}\sqrt{h}\\
&=-\frac{1}{16\pi G}\int_{\varepsilon}^{\Lambda}\mathrm{d}r\int_{\mathbb{T}^2_t}\mathrm{d}x \, \mathrm{d}y\,\frac{(-4)}{r^3}-\frac{1}{8\pi G}\int_{{\mathbb{T}^2_t}}\frac{\mathrm{d}x \, \mathrm{d}y}{\varepsilon^2}+\frac{1}{8\pi G}\int_{{\mathbb{T}^2_t}}\frac{\mathrm{d}x \, \mathrm{d}y}{\Lambda^2}\,,
\end{split}
\end{equation}
which vanishes identically upon performing the $r$ integral in the bulk term.

\subsubsection*{Thermal AdS$_{\boldsymbol 3}$}

\begin{figure}
\centering
\begin{tikzpicture}
\fill[gray, opacity = 0.2] (0,0) [partial ellipse = 0:180:3 and 3];
\fill[white] (0,0) [partial ellipse = 0:180:1 and 1];
\draw[thick, latex-latex] (-4.5,0) -- (4.5,0);
\draw[thick, -latex] (0,0) -- (0,4);
\draw[thick] (0,0) [partial ellipse = 0:180:1 and 1];
\draw[thick] (0,0) [partial ellipse = 0:180:3 and 3];
\node[below] at (-3,0) {\small$|\gamma|=1$};
\node[below] at (1,0) {\small$|\gamma|=|x|$};
\node[above] at (4.5,0) {$(\gamma,\bar\gamma)$};
\node[above right] at (0,4) {$r$};
\end{tikzpicture}
\caption{The fundamental domain of thermal $\text{AdS}_3$, defined by the quotient $(r,\gamma,\bar\gamma)\sim(|x|r,x\gamma,\bar{x}\bar{\gamma})$ with $x=e^{2\pi it}$.}
\label{fig:thermal-ads}
\end{figure}
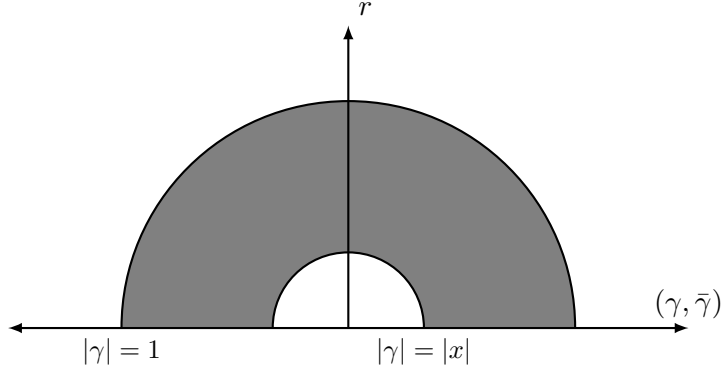

For the case of thermal $\text{AdS}_3$, we can compute the on-shell Einstein-Hilbert action as follows. We again use the metric (using units with $L_{\text{AdS}_3}=1$)
\begin{equation}
\mathrm{d}s^2=\frac{\mathrm{d}r^2+\mathrm{d}\gamma\,\mathrm{d}\bar\gamma}{r^2}
\end{equation}
with the identification $(r,\gamma,\bar\gamma)\sim(|x|r,x\gamma,\bar{x}\bar{\gamma})$. A convenient fundamental domain for thermal $\text{AdS}_3$ is the set of points such that
\begin{equation}
|x|^2<r^2+|\gamma|^2\leq 1\,,
\end{equation}
see Figure \ref{fig:thermal-ads}. With this in mind, it is convenient to choose coordinates
\begin{equation}
r=\ell\sin{\phi}\,,\quad\gamma=\ell\cos{\phi}\,e^{i\theta}\,,\quad\bar\gamma=\ell\cos{\phi}\,e^{-i\theta}\,,
\end{equation}
such that the fundamental domain corresponds to $|x|<\ell\leq 1$. As the divergence in the volume arises as $\phi\to 0$, we can regulate by cutting off $\phi>\varepsilon$. The regularized EH action is
\begin{equation}
I(\varepsilon)=-\frac{1}{16\pi G}\int_{M}\mathrm{d}^3x\sqrt{g}(R+2)+\frac{1}{8\pi G}\int_{\partial M}\mathrm{d}^2x\sqrt{h}K+\frac{1}{8\pi G}\int_{\partial M}\mathrm{d}^2x\sqrt{h}\,.
\end{equation}
The extrinsic curvature of the boundary is given by
\begin{equation}
K=-\frac{1+\cos^2\varepsilon}{\cos\varepsilon}
\end{equation}
and the various terms equate to
\begin{equation}
\begin{split}
-\frac{1}{16\pi G}\int_{M}\mathrm{d}^3x\sqrt{g}(R+2)&=-\frac{\pi t_2}{2G}\left(1-\frac{1}{\sin^2{\varepsilon}}\right)\,,\\
\frac{1}{8\pi G}\int_{\partial M}\mathrm{d}^2x\sqrt{h}K&=-\frac{\pi t_2}{2G}\frac{1+\cos^2\varepsilon}{\sin^2\varepsilon}\,,\\
\frac{1}{8\pi G}\int\mathrm{d}^2x\sqrt{h}&=\frac{\pi t_2\cos{\varepsilon}}{2G\sin^2{\varepsilon}}\,.
\end{split}
\end{equation}
Thus, the regularized EH action is
\begin{equation}
I(\varepsilon)=\frac{\pi t_2}{2G}\left(\frac{1-\cos\varepsilon}{\sin^2\varepsilon}\right)\sim-\frac{\pi t_2}{4G}+\cdots
\end{equation}
To write this in terms of dual CFT data, we can use the Brown-Henneaux formula $G=3L_{\text{AdS}}/2 \mathtt c$ for $L_{\text{AdS}}=1$ to write
\begin{equation}
I_{\text{on-shell}}=-\frac{\mathtt c\pi t_2}{6}\,,
\end{equation}
so that the sphere partition function reads
\begin{equation}
Z_{\text{sphere}}=e^{-I_{\text{on-shell}}}=e^{\frac{\mathtt c\pi t_2}{6}}=|x|^{\frac{\mathtt c}{12}}\,.
\end{equation}
This is in line with the expectation that the dual CFT ground-state energy is $-\frac{\mathtt c}{12}$.

\bibliography{bib.bib}
\bibliographystyle{JHEP.bst}

\end{document}